\documentclass[aps,reprint,superscriptaddress, nobalancelastpage, raggedbottom]{revtex4-2}

 \makeatletter
\def\maketitle{
\@author@finish
\title@column\titleblock@produce
\suppressfloats[t]}
\makeatother
 \usepackage{ragged2e}
\usepackage{amsmath}
\usepackage{amssymb}
\usepackage{physics}
\usepackage[normalem]{ulem}
\usepackage{graphicx}%
 \usepackage{setspace}
 \usepackage[skip=5pt]{caption}
\usepackage{dcolumn}
\usepackage{bm}
\usepackage[dvipsnames]{xcolor}

\begin{document}


\title{Quantum Stabilization and Flat Hydrogen-based Bands \\ of Nitrogen-doped Lutetium Hydride}


\author{Adam Denchfield}
\affiliation{Department of Physics, University of Illinois Chicago, Chicago, Illinois 60607, USA}
\author{Francesco Belli}
\affiliation{Department of Chemistry, University at Buffalo, the State University of New York, Buffalo, New York, 14260, USA}
\author{Eva Zurek}
\affiliation{Department of Chemistry, University at Buffalo, the State University of New York, Buffalo, New York, 14260, USA}
\author{Hyowon Park}
\affiliation{Department of Physics, University of Illinois Chicago, Chicago, Illinois 60607, USA}
\affiliation{Materials Science Division, Argonne National Laboratory, Lemont, Illinois 60439, USA}
\author{Russell J. Hemley}
\affiliation{Department of Physics, University of Illinois Chicago, Chicago, Illinois 60607, USA}
\affiliation{Department of Chemistry, University of Illinois Chicago, Chicago, Illinois 60607, USA}
\affiliation{Department of Earth and Environmental Sciences, University of Illinois Chicago, Chicago, Illinois 60607, USA} 



\date{June 23rd, 2024}

\begin{abstract}
  We explore  electronic and structural properties of  Fm$\overline{3}$m Lu-H-N structures with specific N,H ordering as plausible candidates for near-ambient superconductivity possibly originating from their remarkably narrow hydrogen-based bands at the Fermi level. Although LuH$_{2.875}$N$_{0.125}$ exhibits an instability persisting up to 17 GPa, it is anharmonically stable near ambient pressure when accounting for quantum fluctuations. The presence of flat bands near $E_\text{F}$ is understood to arise from destructive\ quantum interference  between N-p and surrounding H-s orbitals, with certain types of defects leaving the flat bands unaffected.  The results suggest there is an optimal pressure near ambient where the superconducting $T_{\text{c}}$ is maximized  in this structure  by anharmonically-stabilized low-frequency and non-adiabatically coupled high-frequency hydrogen modes.  Despite the metastability of this structure, its electronic properties and dynamical stability when calculated beyond a classical harmonic approach can explain the reported near-ambient superconductivity in Lu-H-N.
\end{abstract}


\maketitle


\section{Introduction}
\label{sec:intro}

The theoretical predictions of high $T_\text{c}$ superconductivity in hydrogen-based metals at megabar pressures \cite{ashcroft1968metallic, ashcroft2004hydrogen} have been realized \cite{somayazulu2019evidence, drozdov2019superconductivity, kong2021superconductivity, troyan2021anomalous} (see Ref. \cite{hilleke2022tuning} for a review), alongside theoretical work studying the chemical precompression effects of hydrogen combined with other elements \cite{pickett2023colloquium}. Notable compounds in this regard are LaH$_{10}$ \cite{liu2017potential, peng2017hydrogen} which was found to agree closely with experiment \cite{somayazulu2019evidence, drozdov2019superconductivity}, followed by later experiments on the YH$_n$ system \cite{kong2021superconductivity, troyan2021anomalous} and ternary hydrides \cite{semenok2021superconductivity, song2023stoichiometric}. 

\begin{figure}[t]
   \centering
     \includegraphics[width=0.99\linewidth]{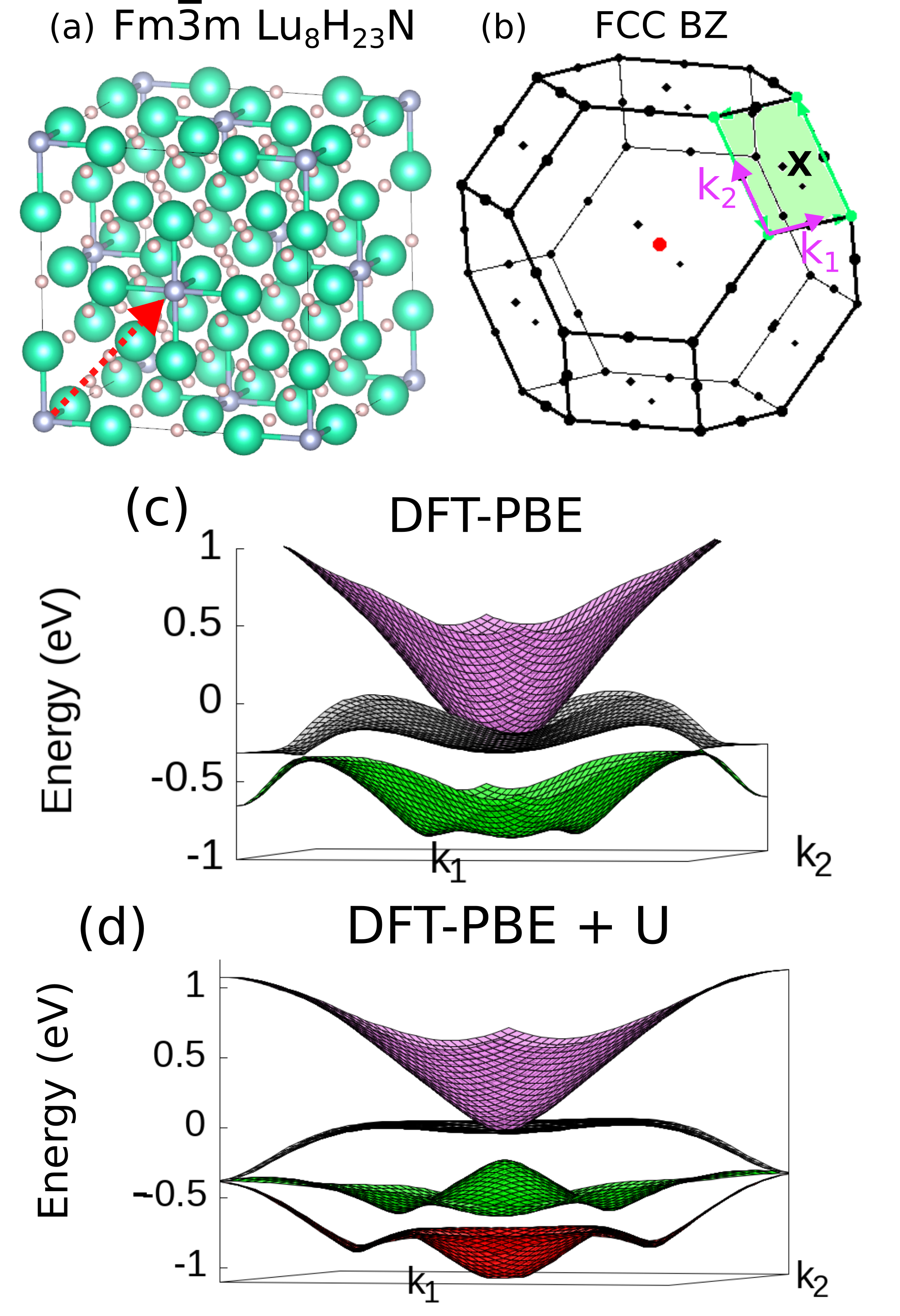}
   \caption{\justifying Crystal structure and band structure of the prototypical Fm$\overline{3}$m structure considered in this paper. (a) Fm$\overline{3}$m LuH$_{2.875}$N$_{0.125}$ structure drawn as an Lu$_8$H$_{23}$N supercell. Atoms are hydrogen (white), lutetium (green), nitrogen (gray). The red arrow indicates the real space direction corresponding to the X point k-vector. (b) Face-centered-cubic (FCC) Brillouin zone with the plane centered on the X-point highlighted as the basis for the band structure plot. (c) 2-dimensional band structure centered on the X-point using DFT and (d) DFT+U. Colors differentiate the bands.}
   \label{fig:Lu8H23N_flatband}
 \end{figure}

Experimental evidence for superconductivity near ambient conditions in the Lu-H-N system has been recently reported \cite{dasenbrock2023evidence, salke2023evidence}. These reports  have  inspired a great deal of computational studies, most of which yielded negative results for supporting near-ambient superconductivity  for the compositions considered  \cite{hao2023first, hilleke2023structure, ferreira2023search, lu2023electron, sun2023effect, lucrezi2024temperature}. On the other hand, density-functional perturbation theory calculations have confirmed that nitrogen doping of LuH$_3$ increases the electron-phonon coupling (EPC) up to over $\lambda > 3$ (arising from multiple modes)  in some structures \cite{hao2023first, fang2023assessing}. Specifically,  Pm$\overline{3}$m LuH$_{2.75}$N$_{0.25}$ was predicted to have $\lambda \approx 3$ and  a $T_\text{c}$ of 100 K if quantum stabilization effects are taken into account \cite{fang2023assessing}.


\begin{figure*}
   \centering
     \includegraphics[width=1.0\linewidth]{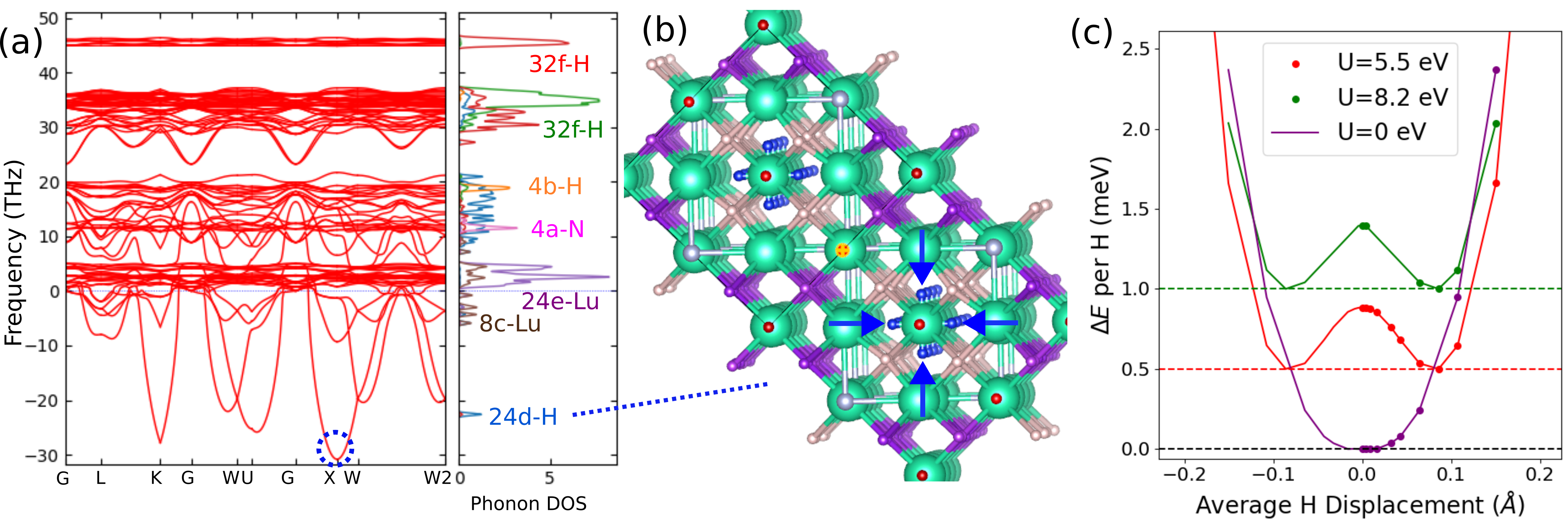}
   \caption{\justifying (a) The phonon dispersion interpolated from a frozen-phonon calculation performed on supercells of the primitive cell (left) and the atom-projected phonon density of states (phDOS). Distinctions between atoms of unique Wyckoff positions are made for clarity. Imaginary frequencies are plotted as negative. The circle highlights that the mode with the largest imaginary frequency is centered on the X-point. (b) Visualization of the structure modulated with the phonon mode with the largest imaginary frequency (blue circle in (a));  32f-hydrogen (pink), 4d-hydrogen (red), the 24d-hydrogen participating in the phonon mode (blue), lutetium (green), nitrogen (gray). Arrows indicate the planar breathing mode of the blue hydrogen atoms. (c) Anharmonicity of the the total DFT energy as a function of mode eigenvector amplitude, yielding a potential energy surface (PES). Points are explicit DFT calculations, with lines connecting each point. The PES was mirrored due to the crystal symmetry, though we did explicitly compute a few inverted points to confirm (not shown).}
   \label{fig:Lu8H23N_phonon_combined}
 \end{figure*}

In contrast to  focusing on the structures on/near the convex hull of thermodynamic stability (e.g. Refs. \cite{hao2023first, hilleke2023structure, ferreira2023search, lu2023electron}), we searched for structures with electronic properties suitable for superconductivity. Following this approach  we reported that a class of LuH$_{2.875-\delta}$N$_{0.125}$ structures supports hydrogen bands at $E_\text{F}$ \cite{denchfield2024electronic}. Furthermore, the representative structure Fm$\overline{3}$m LuH$_{2.875}$N$_{0.125}$ [Fig.\ \ref{fig:Lu8H23N_flatband}(a)] exhibits a 'metallic hydrogen' sublattice whose states form a nearly-flat band 0.2 eV below $E_\text{F}$, with correlation effects raising the band to $E_\text{F}$ [Fig. \ref{fig:Lu8H23N_flatband}(c)]. Subsequent calculations support the flat hydrogen band placement at $E_\text{F}$ \cite{wu2023investigations}.  Two simple estimation methods based on electronic structure indicators and Bardeen-Cooper-Schrieffer (BCS) values relative to similar structures predicted $T_\text{c}$ to be in the range of 165-200 K \cite{denchfield2024electronic}. A key finding in our previous work was the importance of having  N in octahedral positions to produce a large density of hydrogen-based electronic states at the Fermi energy [DOS$_H$($E_\text{F}$)], an observation confirmed by subsequent theoretical studies of Lu-H-N structures with N in octahedral positions \cite{wu2023investigations, pavlov2023anatomy, fang2023assessing, hao2023first}. Moreover, angle-resolved photoemission (ARPES) provided indirect evidence for this or a closely related structure, as it observed a flat band 0.2 eV below the Fermi energy ($E_\text{F}$) \cite{liang2023observation}. 

Here we elucidate in more detail the  dynamical  stability and electronic properties of Fm$\overline{3}$m LuH$_{2.875}$N$_{0.125}$ \cite{denchfield2024electronic} and related structures using density functional theory (DFT) \cite{kohn1965self, giannozzi2009quantum, giannozzi2017advanced} and DFT+U \cite{wang2016local, tolba2018dft+}. Dynamical stability analysis indicates Fm$\overline{3}$m LuH$_{2.875}$N$_{0.125}$ has an anharmonic potential energy surface (PES) at low pressures, but the high-symmetry structure is stabilized by quantum nuclear effects using stochastic self-consistent harmonic approximation (SSCHA)\ calculations \cite{monacelli2021stochastic}. Analyzing the Lowdin charges  of LuH$_3$ and LuH$_{2.875}$N$_{0.125}$, we find that the competition between LuN and LuH$_{3-\delta}$ ionic bonding conspires to form a metallic hydrogen sublattice with conduction states at $E_\text{F}$. This also forms flat bands, which is attributed to local\ quantum interference between N-p orbitals and surrounding H-s orbitals; in other words, localized nitrogen and hydrogen orbitals have alternating hopping matrix elements between them, which leads to a near cancellation of oscillatory factors in the single-particle dispersion $\epsilon_{nk}$ near $E_\text{F}$. We illustrate how the Fm$\overline{3}$m LuH$_{2.875}$N$_{0.125}$ electronic structure (in particular, the flat hydrogen band) is resilient to various defects. We also find that the pressure dependence of the electronic and structural properties is consistent with the superconducting $T_\text{c}$ as a function of pressure suggested for the Lu-H-N system \cite{dasenbrock2023evidence, salke2023evidence}. We conclude by arguing that there is high potential for superconductivity in Fm$\overline{3}$m LuH$_{2.875}$N$_{0.125}$ due to  a combination of non-adiabatic EPC with high-frequency hydrogen modes due to the nearly-flat bands and high EPC from anharmonically stabilized low-frequency hydrogen modes. We discuss the implications of our calculations for synthesis of the material.

\section{Results}
\label{sec:results}

\subsection{Structural Properties of Fm$\overline{3}$m LuH$_{2.875-\delta}$N$_{0.125}$}


\paragraph{Phonon Dispersion.}

 We first address the dynamical stability of Fm$\overline{3}$m LuH$_{2.875}$N$_{0.125}$. We computed the harmonic phonons of the structure relaxed using DFT  without U at various pressures, and observe dynamical instabilities up to 20 GPa. We performed the relaxation and phonon calculation at the DFT-PBE level (see Sec. \ref{sec:methods} for details). The phonon dispersion obtained using frozen-phonon calculations at 10 GPa are shown in Fig. \ref{fig:Lu8H23N_phonon_combined}(a). We examine the phonons at this pressure because DFT calculations at the LDA/PBE level typically overestimate the amount of pressure required to stabilize the FCC RH$_3$ phases by around 10-15 GPa [See Refs. \cite{palasyuk2005pressure, kong2012structural}], elucidated in more detail in a recent SSCHA study of LuH$_3$ \cite{lucrezi2024temperature}. Some Lu modes have small imaginary frequencies, but the modes with strongest imaginary part are hydrogen modes. The mode with the largest imaginary frequency is in the same part of the Brillouin zone as the center of the flat band [compare Figs. 1(b) and 2(a)], indicative of phonon softening due to the flat band.  While a phonon dispersion like this typically indicates a distortion away from the parent structure, we now show this is unlikely to be the case here.
 
 \paragraph{Structural Instability.}
 \label{sec:phon}
 
 The phonon mode with the largest imaginary eigenvalue is visualized in Fig. \ref{fig:Lu8H23N_phonon_combined}(b). In this mode, the 24d (octahedral)  hydrogens (we use the unperturbed Wyckoff position labeling for consistency) move towards/away from the Lu in the 8c positions, with 32f (tetrahedral)  hydrogens moving to occupy space left by the perturbed 24d hydrogens. Such modes in binary hydrides \cite{tanaka2017electron} including many RH$_{3-\delta}$ compounds \cite{kai1989heat, tamatsukuri2023quasielastic, kume2007high, tunghathaithip2016pressure, drulis2009low} exhibit intriguing behavior; for example,  Raman measurements show that a Y-H breathing mode in YH$_{3-\delta}$ is significantly broadened \cite{racu2006strong}, which implies strong electron-phonon coupling  in YH$_{3-\delta}$ and therefore LuH$_{3-\delta}$. As large electron-phonon coupling causes softening of the harmonic DFT phonons in the case of LuH$_{2.875}$N$_{0.125}$, we find  it is important to identify that all observed phonon instabilities only involve the motions of the 8c-Lu, 24d-H, and 32f-H atoms in the unit cell. A close inspection of the contributions to DOS($E_\text{F}$) yields that the 24d and 32f hydrogens contribute to the large DOS($E_\text{F}$) peak (see Fig. \ref{fig:phonPDOS}), whereas the 4b octahedral hydrogens not participating in the phonon mode (colored red) do not contribute to the DOS($E_\text{F}$). This implies that the electron-phonon coupling is strongest for the 24d and 32f hydrogens, which are the hydrogens closest to nitrogen atoms. We note it is the stable 4b hydrogen that are substituted for N to make Pm$\overline{3}$m Lu$_4$H$_{11}$N (LuH$_{2.75}$N$_{0.125}$) \cite{fang2023assessing}.

We also find these planar breathing modes in Pm$\overline{3}$m LuH$_{2.75}$N$_{0.25}$, whose isotropic Eliashberg $T_\text{c}$ was computed to be 100 K with $\lambda \approx 3$ \cite{fang2023assessing}. We do not find this coincidental. In that system, such instabilities were cured at 20 GPa by accounting for quantum and thermal effects via the SSCHA \cite{monacelli2021stochastic}. As the computational time for performing the SSCHA on the Fm$\overline{3}$m LuH$_{2.875}$N$_{0.125}$ cell would be very expensive, we first use the following quasi-classical calculations to investigate whether such a calculation is worth it. 

\paragraph{Phonon Anharmonicity Analysis.}

In order to estimate the energy gain for the distortion along the mode whose frequency has the largest imaginary amplitude, we modulate a 2x1x1 supercell with a distortion corresponding to the corresponding eigenvector of the dynamical matrix, and compute the total DFT energies at various amplitudes of this modulation [Fig. \ref{fig:Lu8H23N_phonon_combined}(c)]. We plot the PES as a function of average hydrogen displacements. We see the slope of energy with respect to distance is extremely small near zero displacement. For U=0, the PES is nearly flat near $x=0$, with the energy of distortion lowered by less than 0.1 meV per hydrogen.  The nearly-flat PES suggests a much smaller imaginary phonon frequency than computed from the \texttt{phonopy} results, which can arise because \texttt{phonopy} only computes the harmonic contribution. The discrepancy between the two approaches is expected when there is sizable anharmonicity (coupling between harmonic eigenmodes), which our present analysis confirms. We also performed the PES calculations with DFT+U  assuming that the relevant correlated subspace are Lu$_d$ orbitals. The results show that the minimum now occurs around at an average displacement of approximately 0.1 \AA, with the DFT+U energy difference between the extrema still corresponding to only 0.4 meV per hydrogen. This difference is still easily small enough for quantum nuclear effects to overcome,  considering the mean-squared displacement $\sqrt{\ev{(\Delta x)^2}}$ for hydrogen from quantum nuclear effects is typically larger than 0.2 \AA\ (e.g. 0.24 \AA\  in LiH \cite{boronat2004quantum}). We also note at lower pressures the PES becomes increasingly anharmonic for all U-values.

\begin{figure}
   \centering
     \includegraphics[width=0.95\linewidth]{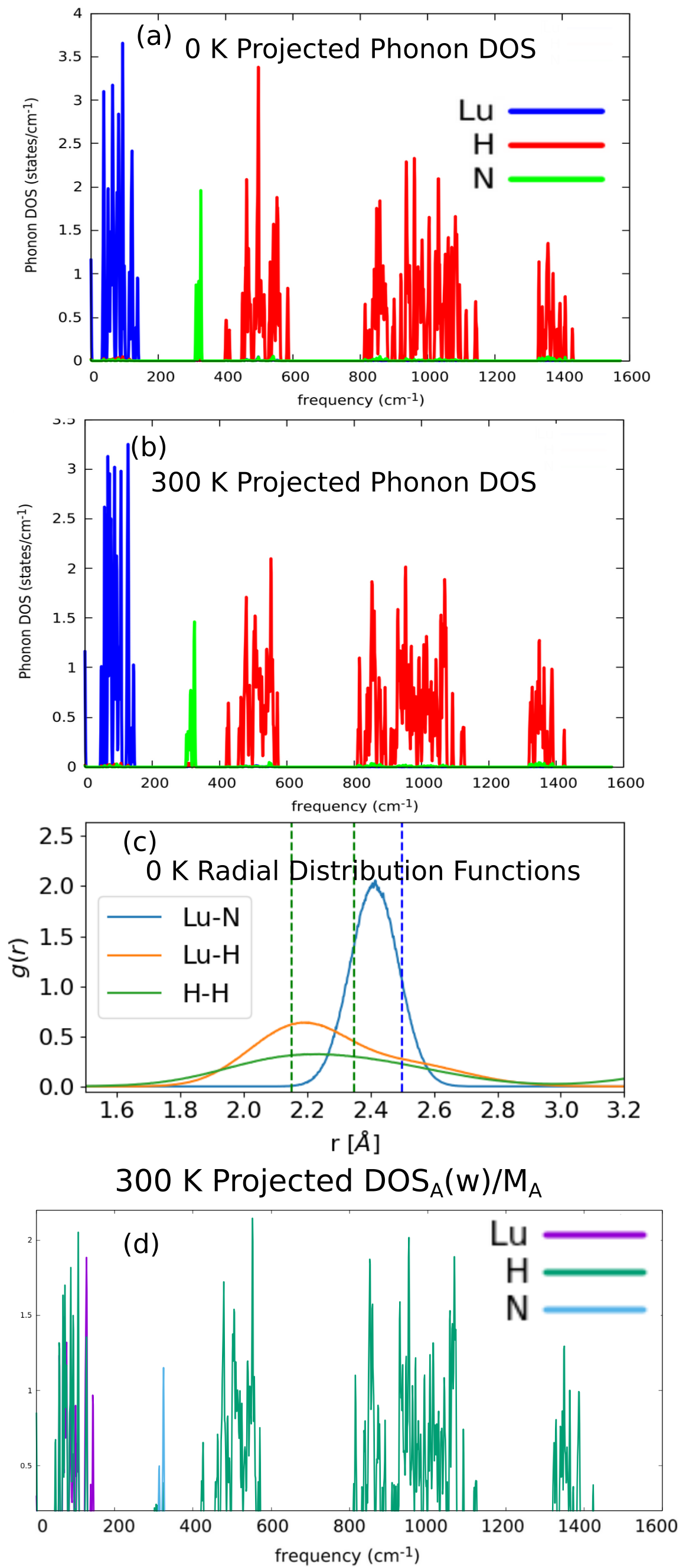}
   \caption{\justifying (a,b) Projected phonon DOS of Fm$\overline{3}$m LuH$_{2.875}$N$_{0.125}$ at 0 K and 300 K computed using the SSCHA accelerated by machine learning. (c) The radial distribution functions for near-neighbor atoms. Green dashed lines indicate classical H-H distances, and the blue dashed line indicates the bond length corresponding to the 24e-Lu being in the symmetric x=0.25 position. (d) The inverse-mass-weighted phDOS. Calculations were performed for $a = 10.04$ \AA\ (ambient pressure for PBE). As quantum nuclear effects generally expand the lattice, the SSCHA calculations predict a slightly higher pressure at this volume of 2.2 GPa.}
   \label{fig:SSCHA_results}
 \end{figure}
 
The collective coordinate analysis shows the PES is clearly not quadratic at low displacements, but is at large displacements, indicating a localized double well of the form $V(x) \sim a x^2 + b e^{-cx^2}$. As was done for SrTiO$_3$ \cite{shin2021quantum, esswein2022ferroelectric}, we solve the Schrodinger equation using the PES in Fig. \ref{fig:Lu8H23N_phonon_combined}(c) and find the ground state density of the hydrogen for this PES is peaked at the center (Fig. \ref{fig:phonModel}).  In comparison, a deeper double well or much heavier atom would have near-degenerate ground and excited states with peaks centered in each well, and the ground state would be susceptible to spontaneous symmetry breaking (see Fig. S3), indicating a structural distortion.  

Analyzing the projected density of states (PDOS) of the PES energy minimum for the case of DFT+U (Fig. \ref{fig:phonPDOS}), we find 24d and 32f hydrogen contributions at $E_\text{F}$ drop to 5-10 meV below $E_\text{F}$, and the 4b hydrogens now contribute a small amount to the DOS($E_\text{F}$) (more per atom than the 24d and 32f hydrogens). A similar change occurs for the U=5.5 eV PDOS. Hence, the  electronic properties of the distorted structure are not particularly sensitive to the exact U-value used. We also note that shifting the 24e-Lu atoms from their relaxed x=0.2419 Wyckoff positions to their higher symmetry x=0.25 positions (with energy cost 9.16 meV/atom for U=5.5 eV) leads to a remarkable increase in the hydrogen DOS($E_\text{F}$) (Fig. \ref{fig:Lu_sym}). 

\begin{figure*}
   \centering
     \includegraphics[width=0.88\linewidth]{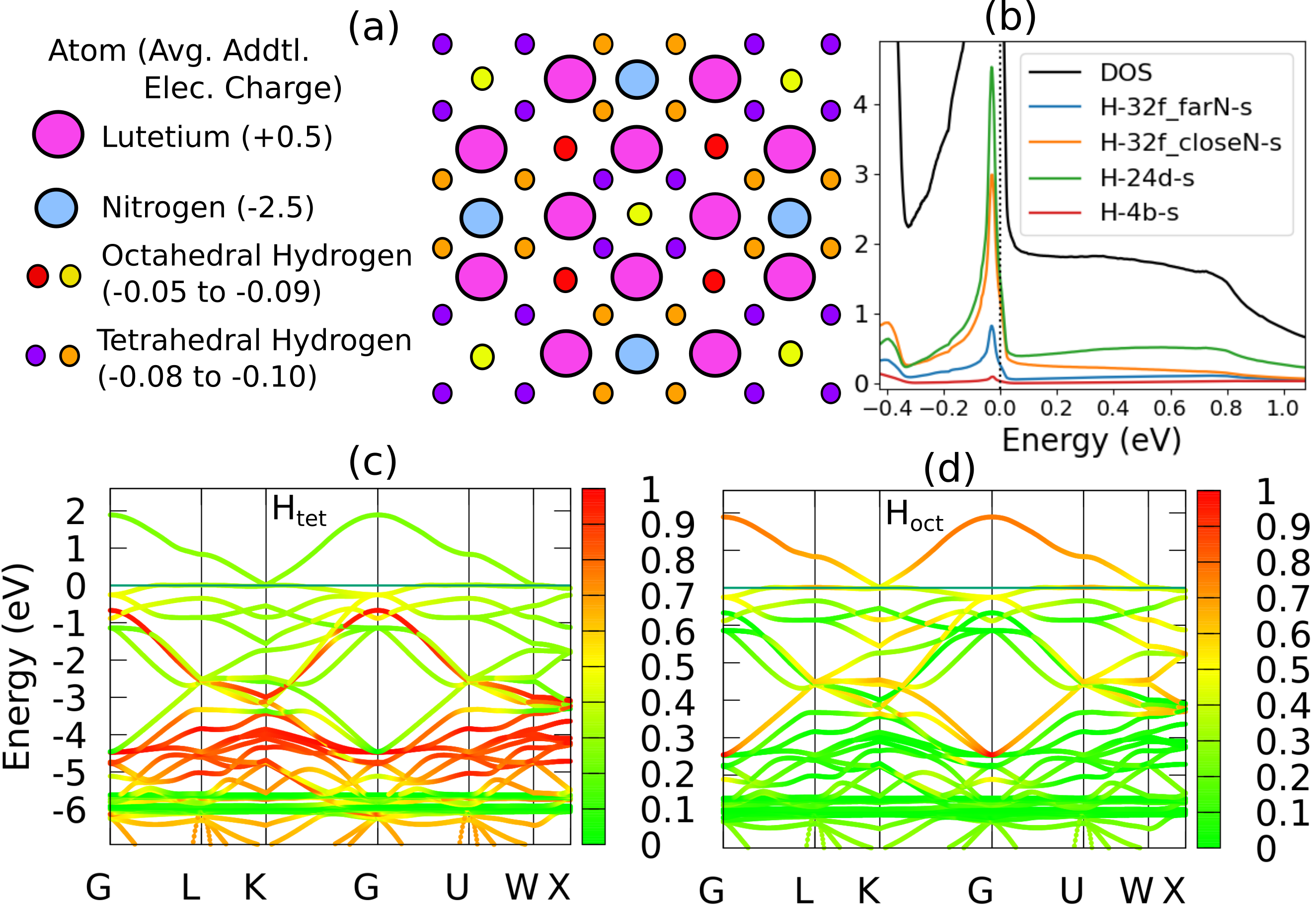}
   \caption{\justifying (a) Lowdin charges (relative to neutral) of each atom in Fm$\overline{3}$m LuH$_{2.875}$N$_{0.125}$, with a schematic layer of the structure. The two different colors for octahedral and tetrahedral hydrogen are because there are two inequivalent H$^{\text{tet}}$ and H$^{\text{oct}}$ Wyckoff positions. (b) The total contributions of the unique hydrogen types to the DOS($E_\text{F}$) peak. (c,d) the wannier projections of the DFT+U band structure for tetrahedral (left) and octahedral (right) hydrogen states, confirming their hybridization at $E_\text{F}$. }
   \label{fig:lowdin}
 \end{figure*}
 
\paragraph{Anharmonicity of Fm$\overline{3}$m LuH$_{2.75}$N$_{0.125}$.}

 We also investigate the effect of vacancies on the electronic and stability properties of Fm$\overline{3}$m LuH$_{2.875-\delta}$N$_{0.125}$. We remove the 4d-H atoms ($\delta = 0.125$) to yield Fm$\overline{3}$m LuH$_{2.75}$N$_{0.125}$, with our calculations discussed in Sec. \ref{sec:Lu8H22N} (also see Fig. \ref{fig:LuH22N_phon}). While the structural instabilities are similar to the $\delta=0$ case, the vacancies make the crystal more prone to a classical structural distortion, which is augmented with the formation of nearly-localized 'hydrogen-cage' states right above $E_{\text{F}}$. It is visually suggestive of metallic superhydride units embedded in the crystal. 

\paragraph{Quantum Nuclear Effects on Stability.}
 
 It has been shown that quantum and thermal effects serve to anharmonically stabilize cubic LuH$_3$ at ambient pressures \cite{lucrezi2024temperature} despite harmonic DFT calculations requiring 20 GPa to stabilize the structure \cite{dangic2023ab}. Similarly, quantum nuclear effects stabilize Pm$\overline{3}$m LuH$_{2.75}$N$_{0.25}$ at 20 GPa \cite{fang2023assessing} whereas harmonic DFT calculations require 60 GPa \cite{hao2023first}. Given these results and the anharmonicity analysis above, we performed SSCHA calculations \cite{monacelli2021stochastic} on Fm$\overline{3}$m LuH$_{2.875}$N$_{0.125}$ (Fig.\ \ref{fig:SSCHA_results}). Simulations were performed for the $\Gamma$-point phonons of the 128-atom conventional cell and made feasible by using machine learning acceleration, as over one million configurations had to be sampled for convergence (see Sec. \ref{sec:methods}). As expected, the quantum corrections stabilize the imaginary modes [Fig. \ref{fig:Lu8H23N_phonon_combined}(a)] with temperature effects providing only small changes. The results indicate that quantum nuclear effects dynamically stabilize this structure, ensuring it is at least metastable. Since SSCHA calculations were performed at the PBE level, the dynamical stability may be slightly overestimated; e.g. Fig.\ \ref{fig:Lu8H23N_phonon_combined}(c) shows correlation effects can enhance the tendency for distortion,  but such effects were not included in the SSCHA.

 The SSCHA-relaxed structure has very minor structural changes from the original DFT-relaxed structure \cite{denchfield2024electronic} and produces the same band structure. The radial distributions of near-neighbor distances \cite{monacelli2021stochastic} at 0 K are plotted in Fig. \ref{fig:SSCHA_results}(c). The nearest-neighbor H-H distance is very broad due to the quantum nuclear effects. The classical DFT-level H$^{\text{tet}}$-H$^{\text{oct}}$ and H$^{\text{tet}}$-H$^{\text{tet}}$ bond lengths are 2.14 \AA \ and 2.35 \AA \ respectively. Temperature effects broaden the Lu-N bond length, which may have significant effects on the DOS($E_\text{F}$) (see Fig. \ref{fig:Lu_sym}).

 To estimate how the phonon modes may contribute to the EPC, we also plot the atom-projected phonon DOS weighted by the inverse atomic weight [Fig. \ref{fig:SSCHA_results}(d)], which gives a rough idea of the hypothetical $\alpha^2 F (\omega)$. We find a significant contribution from hydrogen vibrations at low frequencies, which can barely be seen in Fig. \ref{fig:SSCHA_results}(a). If the EPC is mainly derived from hydrogen-based modes, then this sketch of $\alpha^2 F(\omega)$ is quite consistent with the $\alpha^2 F(\omega)$ computed for LuH$_{2.75}$N$_{0.125}$ \cite{fang2023assessing}. Some relevant equations for Fig. 3 are given in Sec. \ref{sec:fig3}, and one can see that the inverse mass weighted phDOS gives an indication of the contributions to $\alpha^2 F(\omega)$.

Further investigation of the classical-nuclei distortions of LuH$_{2.875}$N$_{0.125}$ indicate that the structure is susceptible to the formation of hydrogen trimers (with H-H distances of 1 \AA) and 7-hydrogen complexes with H-H distances 1.6-1.9 \AA\ weakly coupled by Lu$_d$ orbitals (Sec. \ref{sec:distort}). The DFT energy saved is over 22 meV/atom, showing that the quantum nuclear effects stabilizing the high-symmetry structure are quite strong. This illustrates the competition between hydrogen molecularization and the metallic (possibly superconducting) state of LuH$_{2.875}$N$_{0.125}$, similar to that predicted for LaH$_{10}$\cite{van2023competition}. 


%
   \begin{figure*}
   \centering
     \includegraphics[width=1.0\linewidth]{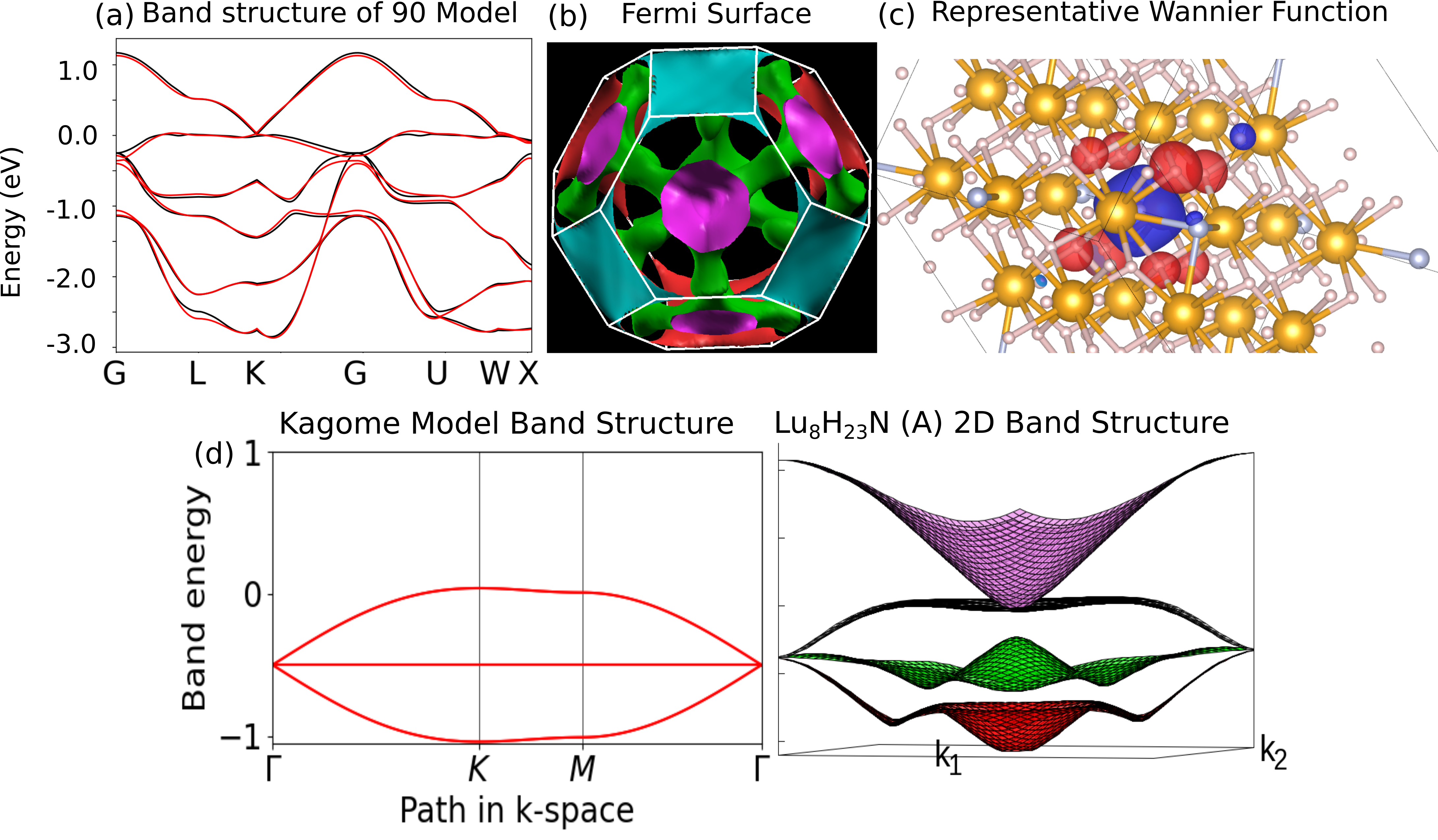}
     \caption{\justifying (a) Band structure comparison between the 6-orbital Wannier90 tight-binding model with all hoppings included (black) vs when only hoppings below a distance of 12 \AA\ are included (red). (b) Associated Fermi surface of the 6-orbital model in agreement with DFT calculations \cite{denchfield2024electronic}. (c) Representative Wannier function, showing an extended H$^{\text{oct}}$ s-orbital surrounded by 8 H$^{\text{tet}}$ s-orbitals. (d) Comparison between a simple 3-orbital Kagome model (left) and the bands of Fm$\overline{3}$m LuH$_{2.875}$N$_{0.125}$ near $E_\text{F}$ centered on the X-point (right). The tight-binding model parameters are given in Fig. \ref{fig:kagome}. }
   \label{fig:WF_cartoon}
 \end{figure*}

\subsection{Ionic Bonding and the Flat Band at $E_\text{F}$}
\label{sec:flat}

\paragraph{Lowdin Charge Analysis.}

We now examine in more detail why the nitrogen/hydrogen states at $E_\text{F}$ give rise to the previously described flat band in Fm$\overline{3}$m LuH$_{2.875}$N$_{0.125}$ \cite{denchfield2024electronic}. While it is difficult to interpret absolute measures of charge via DFT calculations, we can estimate charge transfer effects with respect to a reference system. In our case, we use FCC LuH$_3$ as a reference and compute the Lowdin charges of Lu atoms to be 23.8 (Lu$^{1.2+}$) (not counting core electrons) and the Lowdin charges of H$^{\text{oct}}$ and H$^{\text{tet}}$ to be 1.41 (H$^{0.4-}$). While these values cannot be taken too seriously, we note IR measurements estimate the hydrogen charges in the similar system YH$_{3-\delta}$ to be H$^{0.5-}$ \cite{rode2001evidence}. In comparison, the Lowdin charges of Fm$\overline{3}$m LuH$_{2.875}$N$_{0.125}$ have Lu$^{0.5+}$, N$^{2.5-}$, and H$^{0.05/0.10-}$ [Fig. \ref{fig:lowdin}(a)]. This illustrates the effect of nitrogen's high electronegativity  and preferential charge transfer from Lu, resulting in a more pronounced ionic bond than with the hydrogens. Interestingly, the charge on the hydrogens is close to zero, similar to the predictions for high-$T_{\text{c}}$ LaH$_{10}$  \cite{belli2021strong, somayazulu2019evidence, drozdov2019superconductivity, liu2019microscopic}.

An analysis of the PDOS shows the H$^{\text{tet}}$ ions closest to N atoms bring many more states to $E_\text{F}$ than the hydrogen farther from N [Fig. \ref{fig:lowdin}(b)]. They hybridize with the octahedral hydrogens/nitrogens to form a metallic hydrogen-dominated sublattice that gives rise to hydrogen-based flat bands around $E_\text{F}$. The H$^{\text{tet}}$ closest to N atoms form the smaller honeycomb-like structure seen in the top slice of Fig. \ref{fig:kagome}(a), whereas the H$^{\text{tet}}$ farther from N form larger conduction rings with the octahedral hydrogen. The most dominant contribution at $E_\text{F}$ comes from 24d hydrogens that are sandwiched between two nearby N atoms [red hydrogens in Fig. \ref{fig:lowdin}(a)], which can be seen to be critical to the formation of both types of ring-like structures. Importantly, the hydrogen states that contribute significantly to DOS($E_\text{F}$) are those associated with the soft phonon modes, indicating the role of electron phonon coupling. 
 
 We note that this nitrogen-induced charge transfer effect varies with changing U (in context of DFT+U). This effect is not surprising, as U=8.2 eV for the Lu$_d$ orbitals is needed to describe the LuN band gap with our choice of pseudopotentials. The bandgap shifts from 0.2 eV to its experimental value of 1 eV \cite{topsakal2014accurate} with this U value \cite{denchfield2024electronic}; indeed, raising U tends to shift the energy of H$^{\text{tet}}$ states higher relative to $E_\text{F}$ [see also Fig. \ref{fig:Lu8H23N_flatband}(c)]. 
 
 \paragraph{Tight-Binding Model.}
 \label{sec:TB}

 We created tight-binding models from the DFT calculations on the 32-atom primitive cell in order to capture the effective electronic structure near $E_\text{F}$ (see Sec. \ref{sec:methods}). We began with models using H$^{tet}_s$, H$^{oct}_s$, N$_p$, and Lu$_d$ as initial projections and reproduce the DFT band structure in a large ($\pm 5 $ eV) range around $E_\text{F}$. Removing the Lu$_d$ orbitals yields a model with 26 Wannier functions (WFs), which again reproduces the bands in a wide range around $E_\text{F}$. However, a 10-WF model using H$^{oct}_s$ and N$_p$ initial projections failed to reproduce the band structure near $E_\text{F}$ without unreasonably long-range hoppings (40 \AA).

 We use the 26-WF model to understand how quantum interference can give rise to the flat band in LuH$_{2.875}$N$_{0.125}$. Inspired by the ring-like structures shown in Fig. \ref{fig:kagome}, we plot the corresponding nitrogen-hydrogen sublattice showing the nearest-neighbor hopping parameters (colored to indicate sign) in Fig. \ref{fig:WFs_26orb}. A complex triangular lattice is illustrated, where nitrogen hoppings to the nearest H$^{\text{tet}}$ have alternating sign. This is reminiscent of Kagome and Haldane models, which produce flat bands \cite{misumi2017new}, though our supercell structure necessarily requires a more complex model. To illustrate how quantum interference can arise from alternating hoppings, we show how a simple 3-orbital Kagome lattice tight-binding model (for its parameters, see Fig. \ref{fig:kagome}) reproduces the flat bands  centered on the X-point  near $E_\text{F}$ surprisingly well [Fig. \ref{fig:WF_cartoon} (d-e)], though also reproducing the other (non-flat) bands in the rest of the Brillouin zone would require more orbitals.
 
 It is also possible to obtain non-atomic WFs in order to reduce the complexity of the tight-binding model for LuH$_{2.875}$N$_{0.125}$. We use the  selected columns of density matrix  (SCDM) method to define initial projections \cite{damle2018disentanglement} in order to generate extended initial projections, leading to non-atomic WFs. We are able to reproduce the sharp vHs/flat band at $E_\text{F}$ using a total of 6 WFs, though using 7 WFs improves the  nitrogen-based bands below  $E_\text{F}$ considerably. The extended WFs (shown in Fig. \ref{fig:WFs}) have spreads $\Omega_n \approx 8-11$ \AA$^2$ each. The minimization process for the WF spreads was augmented with injection of noise several times during the process to avoid local minima in favor of a global minimum. Our tight-binding model therefore represents a reasonably minimal model of the system's electronic properties. 

 The tight-binding band structure, corresponding Fermi surface, a representative WF, and comparison to a Kagome model are shown in Fig. \ref{fig:WF_cartoon} (see also Figs. \ref{fig:WFs} and \ref{fig:kagome}).  4 WFs out of the 6 WFs  correspond to extended composed of an octahedral hydrogen, the eight surrounding tetrahedral hydrogen, and for some WFs also the nearest H$^{\text{oct}}$ atoms, with the remaining two WFs being nitrogen p-like orbitals somewhat hybridized with nearby tetrahedral hydrogen (Fig.\ \ref{fig:WFs}).  The relatively extended WF size allows us to avoid needing to have a WF for each hydrogen, yielding a minimal model with as few WFs as possible. These extended orbitals are best described as non-bonding or weakly anti-bonding due to the weak increase of the respective single-particle energies with pressure. These nitrogen-based WFs contribute less to the DOS($E_\text{F}$) than those that are hydrogen-based (Fig. \ref{fig:WFs_pdos}). We find that the band structure near $E_\text{F}$ is not particularly sensitive to long-range or small hoppings by comparing the full tight-binding model bands with those from ignoring hoppings below 0.01 eV and longer-range than 12 \AA\ [Fig. \ref{fig:WF_cartoon}(a)]. Interestingly, these WFs are centered on the octahedral H/N atoms, and have alternating-sign NN hoppings as well, suggesting destructive quantum interference in a 3D cubic model can equally be used to interpret the flat bands \cite{leykam2018artificial}.  We list the tight-binding matrix elements in Sec. \ref{sec:tb_elements}. As an example of the alternating signs, we take WFs 1 and 4, noting $H_{1,4}[0,0,-2]=-0.47$, $H_{1,4}[-1,0,-1]=0.46$, $H_{4,1}[1,0,1]=0.46$, and $H_{4,1}[0,0,2]=-0.47$. 

 We note recent observations indicate that emergent flat bands yield observable localized electron states \cite{chen2023visualizing}. Based on our extended WF minimal model, the flat bands in LuH$_{2.875}$N$_{0.125}$ correspond to quasi-localized electrons in the octahedra between Lu atoms, shared between the 8 H$^{\text{tet}}$ and central H$^{\text{oct}}$. Such localized states emergent at $E_\text{F}$ can even cause systems to emulate heavy fermion behavior \cite{ye2024hopping}, including a nonmagnetic 'charge Kondo' resistance anomaly \cite{taraphder1991heavy, matsuura2012theory, checkelsky2024flat}. Such behavior was observed in Tl-doped PbTe where the effect was proposed to be responsible for an anomalously high $T_\text{c}$ \cite{matsushita2005evidence}. Similar hybridization-induced flat bands and resistance anomalies appear to be present in the La$_3$Ni$_2$O$_7$ superconductor \cite{hou2023emergence, cao2024flat}. 
 
\subsection{Flat Band Resilience to Defects}

\begin{figure}
   \centering
     \includegraphics[width=1.0\linewidth]{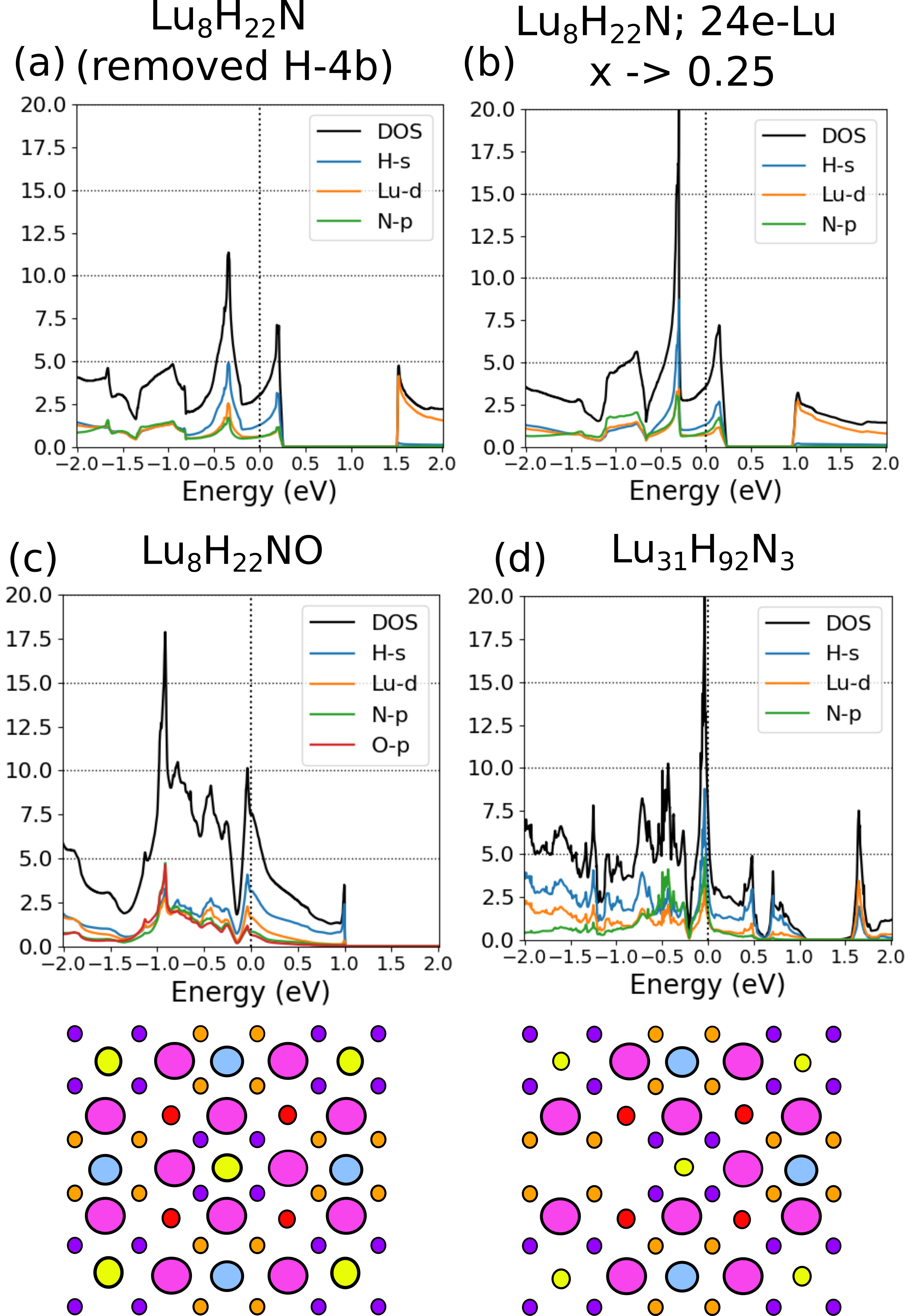}
   \caption{\justifying Projected electronic density of states computed with DFT+U (U on Lu$_d$) for Lu$_8$H$_{22}$N (A) (top), Lu$_8$H$_{22}$NO (middle left), and Lu$_{31}$H$_{92}$N$_3$ (middle right), where a LuN unit was removed from the structure. The drawings at the bottom indicate the modifications of the middle two structures; hydrogen atoms replaced by larger oxygen atoms in the bottom left, and a LuN unit removed from the bottom right. Relaxation was performed only for the top left case.}
   \label{fig:Lu_sym_doping}
 \end{figure}

 We investigate the role of defects on the flat band, extending the hydrogen vacancy analysis we performed previously \cite{denchfield2024electronic}. Removing the 4b hydrogen from Fm$\overline{3}$m LuH$_{2.875}$N$_{0.125}$ only produces an electron-doping effect [Fig. \ref{fig:Lu_sym_doping}(a)]. While the 4b hydrogen were irrelevant to the DOS peak and EPC, they do capture an average of 0.05 electrons. We found that hole-doping the 32-atom primitive unit cell by 1 hole brings $E_\text{F}$ back to the hydrogen peak but otherwise leaves the DOS shape unchanged. This configuration also has the DOS peak height increased by bringing the 24e-Lu to their symmetric x=0.25 positions [Fig.\ \ref{fig:Lu_sym_doping}(b)].

 A more striking perturbation to the system involves replacing the 4b hydrogens entirely with oxygen [Fig.\ \ref{fig:Lu_sym_doping}(c)]. The PDOS reveals a modest decrease of the hydrogen DOS peak compared to Fig. \ref{fig:Lu_sym_doping}(a), but is otherwise unperturbed. Another striking observation is that removing an entire LuN unit from the 128-atom unit cell, (yielding Lu$_{31}$H$_{92}$N$_3$) preserves the flat band [Fig. \ref{fig:Lu_sym_doping}(d)]. These findings suggest that a range of vacancies and structural defects such as grain boundaries do not necessarily destroy the flat bands  under the assumption that the structure does not appreciably distort. 




\begin{figure}
   \centering
     \includegraphics[width=1.0\linewidth]{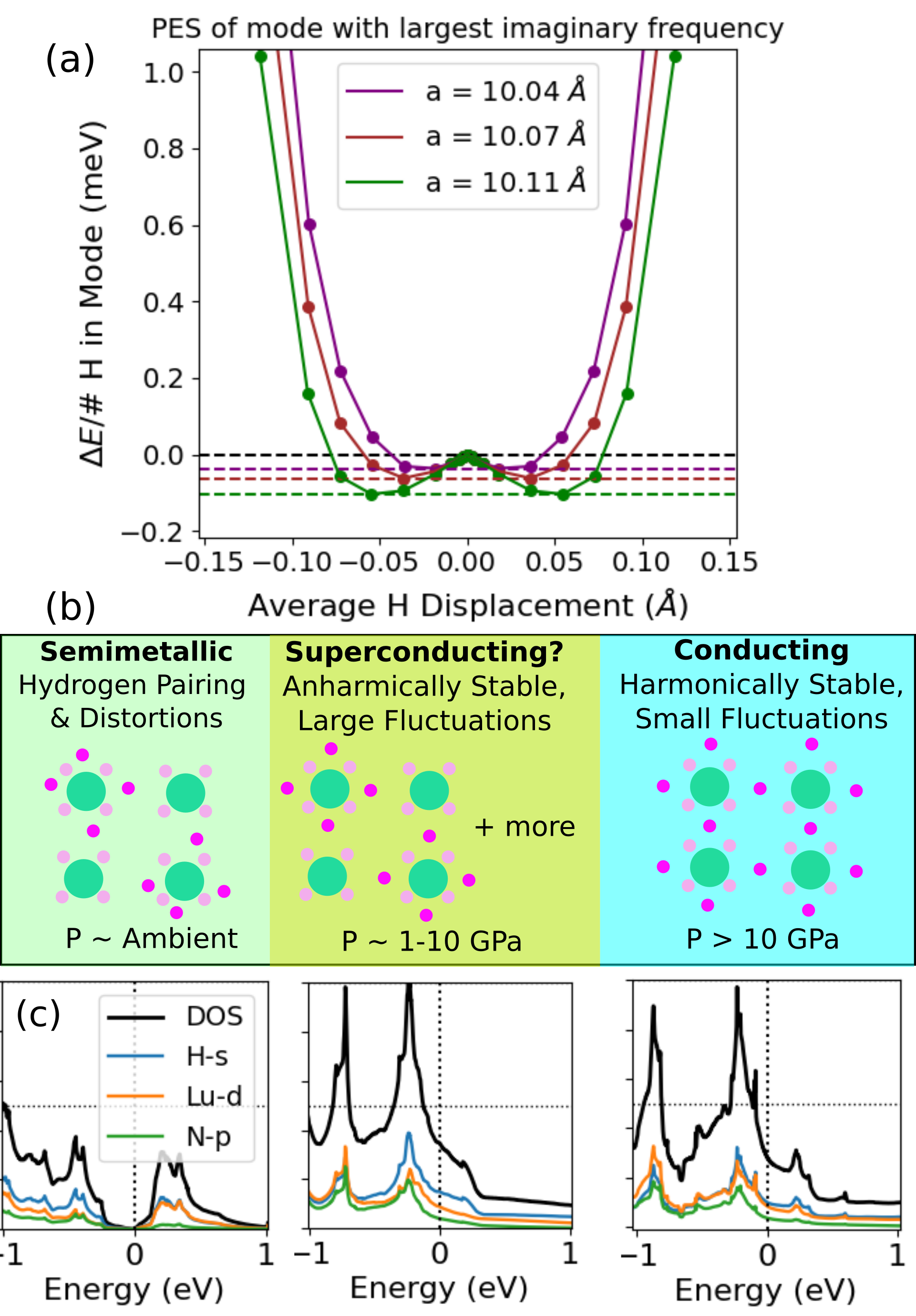}
   \caption{\justifying (a) Pressure dependence of the anharmonic PES of the mode with the largest magnitude of the imaginary frequency  of Fm$\overline{3}$m LuH$_{2.875}$N$_{0.125}$ using the SSCHA-relaxed structure. (b) Schematic sequence of phases suggested by the present calculations. (c) Associated electronic PDOS of Fm$\overline{3}$m LuH$_{2.875}$N$_{0.125}$ in each of the pressure regimes. For high pressure we used the PBE-relaxed structure at 20 GPa. }
   \label{fig:Lu_pressure}
 \end{figure}
 
\section{Discussion}

\subsection{Potential Superconductivity}


We discuss the implications of our findings for potential superconductivity in the Lu-H-N system. We find that a model structure in this system, Fm$\overline{3}$m LuH$_{2.875}$N$_{0.125}$, is dynamically stabilized with the inclusion of quantum nuclear effects. This does not mean that the electron-phonon coupling is small, despite the conventional wisdom that large EPC causes structural distortions \cite{allen1975superconductivity}. In this case, model calculations indicate that when vibrational degrees of freedom greatly outnumber electronic degrees of freedom, CDW distortions can be prevented due to quantum nuclear effects \cite{hofmann2022heuristic}. This is indeed satisfied in Fm$\overline{3}$m LuH$_{2.875}$N$_{0.125}$, whose electronic degrees of freedom near $E_\text{F}$ are primarily composed of simple H$_s$ orbitals. We also find magnetic instabilities are unlikely \cite{denchfield2024electronic} (also see Ref. \cite{ojajarvi2018competition}). The remaining instability is therefore superconductivity.

Unfortunately, full Migdal-Eliashberg calculations for the superconducting gap/$T_\text{c}$ are beyond our computational resources. However, we are able to estimate $\lambda$ using finite differences for q=0 phonons by the method outlined in Ref. \cite{sun2022electron}. The result gives a total $\lambda = 1.81$ despite lacking the contributions from modes with imaginary frequencies. For comparison, a full calculation for FCC LuH$_3$ including quantum/thermal stabilization has $\lambda \approx 1.6$, leading to a $T_\text{c}$ estimate of 50 K \cite{lucrezi2024temperature}. To get an idea of what $\lambda$ should be for the anharmonically stabilized high-symmetry phase of Lu$_{2.875}$N$_{0.125}$, we note SSCHA-stabilized LuH$_{2.75}$N$_{0.25}$ has $\lambda \sim 3$ at 20 GPa \cite{fang2023assessing}, and our own calculations on LuH$_{2.75}$O$_{0.25}$ at 25 GPa also compute $\lambda \sim 3$ with an Eliashberg $T_{\text{c}}$ of 125 K \cite{denchfield2024candidate}. 

Such large $\lambda \geq 2$ brings us to the strong coupling limit  of the Eliashberg equations  \cite{allen1975superconductivity}. In the strong coupling limit T$_c \sim \omega_{\text{ln}}\sqrt{\lambda}$, there is a balance between $\lambda$ and $\omega_{\text{ln}}$; as $\lambda$ increases, it softens phonon frequencies, decreasing $\omega_{\text{ln}}$ \cite{allen1975superconductivity, quan2019compressed}. However, for these hydrides the quantum nuclear effects of hydrogen stabilizes the low frequency modes,  where strong electron-phonon couplings exist. This implies there is a pressure range of a few GPa where quantum nuclear effects keep $\omega_{\text{ln}}$ roughly fixed even as N doping increases $\lambda$  through increasing DOS$_{\text{H}}(E_{\text{F}})$ and increased charge neutrality of hydrogen atoms. 

\subsection{Pressure Dependence of Structure}
\label{sec:press}
The poor treatment of  correlation effects in DFT may lead to an underestimate of the lattice constant of LuH$_{2.875}$N$_{0.125}$; FCC LuH$_{3-\delta}$ has a lattice constant of 5.12-5.16 \AA\ \cite{palasyuk2005pressure, moulding2023pressure} but even DFT-PBE, which is known to overestimate volumes \cite{haas2009calculation}, predicts a much smaller value of 4.99-5.02 \AA\ \cite{liu2023parent, hao2023first}. Our preliminary self-consistent DFT+U calculations within the linear response constrained DFT formalism \cite{cococcioni2005linear, timrov2021self} also indicate that the lattice constant of LuH$_{2.875}$N$_{0.125}$ is expanded by upwards of 1.5\% relative to the PBE values from correlation effects. Therefore, we consider the effect of lattice expansion on the anharmonicity analysis performed previously [Fig. \ref{fig:Lu8H23N_phonon_combined}(c)]. We repeat analyzing the leading structural instability from the phonon dispersion for different lattice constants, using the SSCHA-relaxed structure as a starting point. We find that the depth of the double well (and thus tendency for distortion) increases as the lattice constant is increased [Fig. \ref{fig:Lu_pressure}(a)]. This result is consistent with experiments on other RH$_3$ compounds, which suggest the PES shifts from anharmonic, to flat, to harmonic on compression \cite{kume2007high, kume2011high, tamatsukuri2023quasielastic} and heating \cite{fischer1978neutron, ito1983phase, kai1989heat, majer1999model}. 

Based on estimates from our first-principles calculations, we hypothesize the phase diagram of LuH$_{2.875-\delta}$N$_{0.125}$ takes the form shown in Figure \ref{fig:Lu_pressure}(b). For the ambient pressure phase, we assume that the correlation effects and larger lattice constant both increase the likelihood of distortion at ambient pressures, which takes the form of a  hydrogen molecularization transition  which is semimetallic or semiconducting when looking at the leading instability [Fig. \ref{fig:Lu_pressure}(c); see also Fig. \ref{fig:multicenter}] at low temperatures. At high pressures the hydrogen atoms are all harmonically stabilized, but a large remnant EPC may still result in a high temperature superconductor. At pressures between these two limits (e.g. 1-10 GPa) there is a competition between molecularization and metallicity similar to LaH$_{10}$ \cite{van2023competition}, except in this case the quantum nuclear effects stabilize the high symmetry metallic structure at near-ambient pressures. As the anharmonic PES results in large hydrogen displacements, the associated electron-phonon coupling is likely nontrivially affected, like PdH \cite{errea2013first} and the 'rattling superconductor' KOs$_2$O$_6$ \cite{chang2009cooper, ogusu2010superconducting}. It is possible that the anharmonicity from the proximity to this structural transition would maximize T$_{\text{c}}$, similar to how bulk BaPb$_{1-x}$Bi$_x$O$_3$ exhibits both maximal $T_\text{c}$ and granular superconducting behavior as it approaches its insulating phase \cite{parra2021signatures}.


The electronic structure of Fm$\overline{3}$m LuH$_{2.875-\delta}$N$_{0.125}$ at different pressure ranges illustrates that the flat bands near $E_{\text{F}}$ only exist in the high-symmetry structure. The large effective electron masses associated with the flat bands, in combination with the light mass and nearly-flat PES for the octahedral hydrogen, imply anomalously large vertex corrections to the Migdal-Eliashberg approach \cite{grimaldi1995nonadiabatic, hague2008breakdown}. We note vertex corrections give around a 20\% increase in $T_\text{c}$ for SH$_3$ \cite{schrodi2020full} and FeSe/SrTiO$_3$ interfaces \cite{liu2021towards}. However, vertex corrections can also reduce $T_{\text{c}}$, so we cannot with certainty state $T_{\text{c}}$ in this system would be maximized in this intermediate pressure regime. Our conjectured phase diagram would explain the pressure-dependence of $T_{\text{c}}$ reported in Lu-H-N samples \cite{dasenbrock2023evidence, salke2023evidence}. More precise quantitative results on the pressure ranges of this phase diagram as well as the superconducting properties will require systematic, self-consistent treatment of correlation and quantum nuclear effects, which we plan to pursue in future work. 

\subsection{Role of Nitrogen}

Our results indicate that nitrogen in LuH$_{3-\delta}$N$_\epsilon$ can play a role beyond hole-doping. As Lu preferentially forms stronger ionic bonds with N over H, the hydrogen sublattice can form an extended metallic network at $E_\text{F}$, which remains metastable for this nitrogen doping level. For Fm$\overline{3}$m LuH$_{2.875-x}$N$_{0.125}$, the alternating hoppings between N and nearby hydrogen induces nearly flat bands near $E_\text{F}$, with the highest-energy band (at $E_\text{F}$) having mostly hydrogen character. These effects explain why the $T_\text{c}$ estimates for this system are much higher than that for VCA calculations of hole-doped LuH$_3$ \cite{ouyang2023superconductivity}. While hydrogen orbitals are normally either highly dispersive s-orbital states or fully bonded and nonmetallic, the model structures presented produce metallic hydrogen states with large effective masses whose single-particle Wannier functions are composed of extended orbitals involving 7-9 hydrogen s-orbitals at a time. Depending on the details of the correlation effects, the effective masses of the electrons may approach that of the hydrogen nucleus (the proton). 


\subsection{Implications for Synthesis}

Our predicted metastability of Fm$\overline{3}$m LuH$_{2.875}$N$_{0.125}$ has implications for the synthesis and identification of potential superconducting Lu-H-N. If correlation effects and/or a larger lattice constant result in hydrogen distortion such as in Fig. \ref{fig:multicenter}, LuH$_{2.875-\delta}$N$_{0.125}$ may be semimetallic or semiconducting at ambient pressures. We note our stability calculations for Fm$\overline{3}$m LuH$_{2.875}$N$_{0.125}$ at 2 GPa exhibits stiffer phonons less sensitive to pressure or temperature corrections, features that suggest more resistance toward decomposition than FCC LuH$_3$ \cite{lucrezi2024temperature}. This means we can draw comparisons to the experimental work on cubic LuH$_{3}$. While pristine LuH$_{3-\delta}$ transforms to a hexagonal structure below 12 GPa \cite{palasyuk2005pressure}, a supercell distortion of cubic LuH$_{3-\delta}$ is produced on compression of hexagonal LuH$_3$ to 2 GPa in a nitrogen environment \cite{moulding2023pressure}. This observation is reasonably consistent with SSCHA calculations \cite{lucrezi2024temperature} predicting cubic LuH$_3$ is stable near ambient conditions. Methods used for stabilizing cubic YH$_3$ may also prove useful \cite{kataoka2018stabilization, kataoka2021face, kataoka2022origin}. As hole-doping RH$_3$ typically leads to hydrogen loss in ambient conditions \cite{kataoka2018stabilization}, bulk LuH$_{2.875}$N$_{0.125}$ may require a hydrogen gas environment \cite{kataoka2021face} or a surface coating with an H$_2$ resistant surface film to prevent hydrogen loss. Interestingly, LuH$_3$ itself may serve as a hydrogen diffusion barrier like YH$_3$ \cite{den1998visualization}, which may simplify the synthesis. 

Despite its dynamical stability when including quantum nuclear effects, DFT calculations indicate Fm$\overline{3}$m LuH$_{2.875}$N$_{0.125}$ at ambient pressures is 150 meV/atom above the convex hull \cite{sun2023effect, hao2023first} and therefore may need to be first synthesized at higher pressures or other processes. 
Previous work on nitridization properties of LuH$_3$ may prove useful for the successful synthesis of LuH$_{2.875-\delta}$N$_\epsilon$ structures with N in octahedral positions \cite{dierkes2017metals}. Recently, N-doped FCC LuH$_3$ with approximate stoichiometry LuH$_{2.83}$N$_{0.17}$ was synthesized and recovered to ambient conditions \cite{li2024stabilization} using one of the stabilization procedures we recommended above \cite{kataoka2022origin}. 

The flat band near $E_\text{F}$ observed in via ARPES measurements of Lu-H-N \cite{liang2023observation} is consistent with our predictions for Fm$\overline{3}$m LuH$_{2.875}$N$_{0.125}$ \cite{denchfield2024electronic}. However, we find flat bands near $E_\text{F}$ can be created in other stoichiometries. One example is Lu$_{16}$H$_{47}$N (see also Ref. \cite{hao2023first}) with an Im3m structure, whose electronic structure is unchanged from hole-doped LuH$_{3}$ aside from a flat nitrogen band just below $E_\text{F}$ (Fig. \ref{fig:Lu16H47N}). A comparison of the above composition to LuH$_{2.875}$N$_{0.125}$ \cite{denchfield2024electronic} and LuH$_{2.75}$N$_{0.25}$ \cite{fang2023assessing} illustrates the strong doping-dependence of the electronic structure of  LuH$_{3-\delta}$N$_\epsilon$.

\section{Conclusions}

This work establishes the quantum nuclear effects on the structural and electronic properties of Fm$\overline{3}$m LuH$_{2.875}$N$_{0.125}$. In particular, we show that the structure is dynamically stable at 2 GPa with the PBE functional and unveil the origin of its nearly-flat hydrogen-based band near $E_\text{F}$. Our calculations indicate structures near this stoichiometry (e.g., LuH$_{2.875-\delta}$N$_{0.125}$) are plausible candidates for a superconducting Lu-H-N phase, with fine tuning of the doping, pressure, and small distortions of the octahedral hydrogen all capable of raising $T_\text{c}$.  Our calculations suggest LuH$_{2.875}$N$_{0.125}$ undergoes a pressure-induced phase transition from a semiconducting or semimetallic molecularized hydrogen state to a metallic (and possibly superconducting) state in the 1-10 GPa pressure range. Despite its dynamical stability with quantum nuclear effects, the structures considered here may not be on the convex hull based on first-principles calculations. Synthesis, annealing, and storage methods may all impact the kinetic stability and determine whether such a material will be metastable enough to take advantage of its utility as a near-ambient superconductor.

This research was supported by the U.S. National Science Foundation (NSF, DMR-2104881; R.H.), the U.S. Department of Energy-National Nuclear Security Administration (DOE-NNSA) through the Chicago/DOE Alliance Center (DE-NA0003975; A.D., E.Z., R.H.), the DOE Office of Science (DE-SC0020340, E.Z., R.H.), and NSF SI2-SSE (Grant 1740112; H.P.)

\section{Methods}
\label{sec:methods}
We performed DFT calculations primarily with the \texttt{Quantum Espresso} software package \cite{giannozzi2017advanced}. The SSSP database \cite{prandini2018precision} was used to vet pseudopotentials for convergence of the computed pressure, phonon frequencies, and formation energies. Accordingly, we chose the pseudopotential for Lu from Ref. \cite{topsakal2014accurate}, and the \texttt{pslibrary} pseudopotentials for N and H \cite{dal2014pseudopotentials}. Each relaxed structure had remaining forces under \texttt{1e-5} Ry/Bohr, and energy differences were under \texttt{1e-5} Ry. We used wavefunction plane-wave cutoffs of 75 Ry for the structures and charge density cutoffs of 300 Ry, though we performed convergence tests with higher values and found no differences.  We generally used tight k-meshes corresponding to 0.13 \AA$^{-1}$, but did convergence tests with more k-points as needed, such as for Fig. \ref{fig:Lu8H23N_phonon_combined}(c). We used \texttt{Quantum Espresso}'s default atomic projections for the PDOS calculations, and used maximally localized Wannier functions \cite{marzari2012maximally, wu2018wanniertools} to create the minimal tight-binding model. The \texttt{phonopy} software \cite{togo2023first} was used to compute the phonon spectrum in Fig. \ref{fig:Lu8H23N_phonon_combined}(a), in which the phonons are computed with the finite displacement method. The shown phonon spectrum is interpolated from a $2\times 2 \times 2$ q-mesh. The \texttt{phonopy} calculation for the harmonic modes was performed using the default displacement distances of 0.0105\AA. Density functional perturbation theory (DFPT) calculations \cite{baroni2001phonons} produced similar results. When using DFT+U \cite{wang2016local, tolba2018dft+} we placed U on Lu$_d$ orbitals. 

The SCF calculations for the SSCHA were performed with Quantum Espresso using the PBE functional and a PAW pseudopotential. The cutoff on the wavefunctions was 80 Ry with a 0.02 Ry smearing and a 5$^3$ k-point mesh. A machine learning active learning procedure through the MLIP package \cite{novikov2020mlip} was used in order to reduce the DFT load. 
The SSCHA relaxation was performed with a fixed cell but allowing the atoms to relax to their equilibrium positions. In the SSCHA \cite{Lorenzo}, the displaced configurations for the various supercells used for the MTP training were generated at 300 K and 0 K. The relaxation was performed by fixing the size and shape of the cell, but allowing the atoms to relax to their equilibrium positions. The MLIP package \cite{novikov2020mlip} was employed to generate an MTP of level 10, with cutoff distances of $0.9<x<9$ \AA. Every structures with extrapolation grade $\gamma$select $> 2$ became a candidate for retraining the potential. 
 
\newpage

\bibliography{A.bib}



 \begin{figure*}[t]
   \centering
   \includegraphics[width=0.9\linewidth]{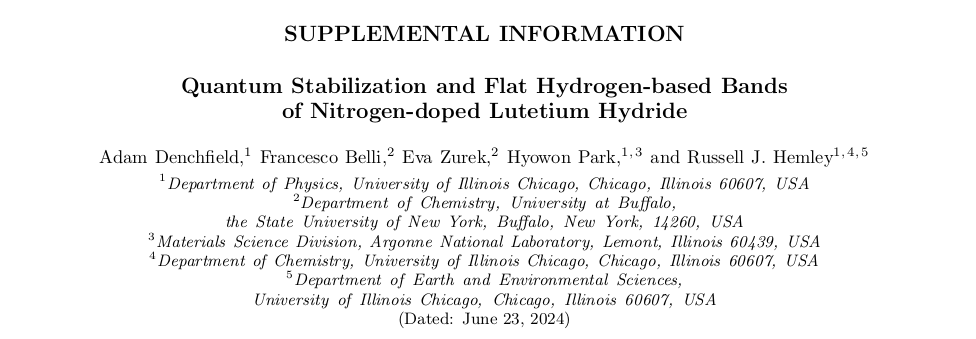}
 \end{figure*}

\clearpage



\renewcommand{\thesection}{SI-\Alph{section}}    
\renewcommand{\thefigure}{S\arabic{figure}}
\setcounter{figure}{0}

\section{T$_c$ Estimate}
\label{sec:tcest}
We performed DFT calculations using \texttt{Quantum Espresso} \cite{giannozzi2017advanced} at the PBE level \cite{perdew1996generalized} to match the calculations used to create the networking-value T$_c$ estimator \cite{belli2021strong}, as using other functionals would invalidate the usage of the estimator. The networking-value T$_c$ estimator \cite{belli2021strong} was derived by correlating the electronic properties and Eliashberg T$_c$s computed for hydride superconductors. It takes the form

 \begin{align}
   \label{eq:tc}
   & T_c \approx 750 \Phi - 85 \quad (\text{Kelvin}) \\
   & \Phi \equiv \phi^{crit}_{net}  \times  H_f \times \sqrt[3]{\text{DOS}_{H,rel}(E_F)}\\
   & \phi^{crit}_{net} \equiv \text{Connectivity of electronic H network} \\
   & H_f \equiv \frac{\text{\# of H forming network}}{\text{\# of atoms in unit cell}}\\
   & \text{DOS}_{H,rel}(E_F)  \equiv \frac{\text{DOS}_H(E_F)}{\text{DOS}_{tot}(E_F)} 
 \end{align} where $\phi^{crit}_{net}$ is the isovalue of the Electron Localization Function (ELF) as computed by \texttt{Quantum Espresso} using nspin=1, where the hydrogen form an extended network. This T$_c$ estimator is reportedly indicative of Eliashberg predicted T$_c$ for hydrides, correct within a range of 60 K \cite{belli2021strong}. For Lu$_8$H$_{23}$N (A) we had computed $\phi^{crit}_{net} = 0.51$, we used H$_f$ = 22/32 since the 4b hydrogen only joined the network at a slightly lower isovalue, and the QE projections resulted in $\text{DOS}_{H,rel}(E_F) \approx 0.5$. This resulted in a networking-value T$_c$ of about 123 K.
 
 In the present work, we used Maximally Localized Wannier Functions (MLWF) \cite{marzari2012maximally} for the atomic projections [see Fig. \ref{fig:lowdin}(c-d)], which are unitary transformations from the Kohn-Sham states, and now find the E$_F$ states are almost entirely hydrogen based. We also simulated 2 GPa of pressure consistent with the experimental conditions \cite{dasenbrock2023evidence}, resulting in an increased $\phi^{crit}_{net} = 0.52$ for H$_f$=22/32. These factors raise the networking-value T$_c$ estimate up to 185 K, becoming consistent with the BCS T$_c$ ratio of 200 K we had mentioned at the end of our previous work \cite{denchfield2024electronic}. This 185 K estimate may be an underestimate since the magnitude of the peak is not accounted for in the networking-value T$_c$ estimator, whereas we clearly have a huge peak. Furthermore, full-bandwidth Eliashberg calculations find for H$_3$S that doping to the DOS peak increases the predicted T$_c$ by a respectable 30 K \cite{lucrezifull}. Therefore, we find it plausible this T$_c$ estimate is a severe underestimation for this structure.

 \section{Anharmonicity of Fm$\overline{3}$m LuH$_{2.75}$N$_{0.125}$}
\label{sec:Lu8H22N}
  We also analyze the structural stability upon introducing some hydrogen vacancies in Fm$\overline{3}$m LuH$_{2.875}$N$_{0.125}$. By removing the 4d-H atoms we obtain LuH$_{2.75}$N$_{0.125}$, which we find experiences similar structural distortions to LuH$_{2.875}$N$_{0.125}$. The PDOS of the high-symmetry structure is also dominated by a hydrogen metallic phase [Fig. \ref{fig:LuH22N_phon}(a)] whose superconducting $T_\text{c}$ is estimated to be 175 K (see Sec. \ref{sec:tcest}). The harmonic $q=0$ phonons are calculated for a 124-atom supercell, with the atom-projected phonon DOS indicating a hydrogen-based distortion as the leading instability [Fig. \ref{fig:LuH22N_phon}(b)]. As before, we investigated the anharmonicity of the total DFT energy with respect to displacing atoms along the mode with the largest imaginary value for its frequency. The computed PES is fairly flat near zero displacement, with minima corresponding to an average hydrogen displacement of 0.14 \AA\ [Fig. \ref{fig:LuH22N_phon}(c)]. The energy gain at these minima is 3.5 meV/hydrogen compared to 0.4 meV/hydrogen in LuH$_{2.875}$N$_{0.125}$, indicating this stoichiometry is more susceptible to these distortions. A similar 1D Schrodinger equation model using the PES in Fig. \ref{fig:LuH22N_phon}(c) for a single hydrogen atom finds the ground state hydrogen nuclear density (inset) is almost flat. Rather than remaining metallic upon distortion along this phonon mode, the structure becomes semiconducting with a PDOS peak just above $E_\text{F}$ [Fig. \ref{fig:LuH22N_phon}(d)]. Given the flat hydrogen ground state density, this implies the system fluctuates between a nearly equal superposition of metallic and insulating states. We visualize the distorted structure [Fig. \ref{fig:LuH22N_phon}(e)], and can link the PDOS peak 0.2 eV above $E_\text{F}$ to quasi-localized states associated with the two Lu-H complexes highlighted in blue. Regarding the isotope effect, the extra mass would concentrate the ground state density at the two minima, resulting in a semiconductor. 
 
\section{Classical Distortions of Fm$\overline{3}$m LuH$_{2.875}$N$_{0.125}$ and Formation of Local 7-H Networks}
\label{sec:distort}

While quantum effects are observed to stabilize the high symmetry metallic phase of Fm$\overline{3}$m LuH$_{2.875}$N$_{0.125}$ at 2 GPa at the PBE level, an examination of the classically preferred distortions is still relevant to understand how the electronic structure changes with quantum fluctuations of the nuclei, as this is associated with the electron-phonon coupling.

To that end, we start with the SSCHA-relaxed structure and let it undergo variable-cell relaxation at 1 GPa. The structure relaxes to form linear hydrogen trimers with internal H-H distances of 1 \AA\ [Fig. \ref{fig:multicenter} (left)]. We also identify two 7-hydrogen complexes in the cell using the Electron Localization Function [Fig. \ref{fig:multicenter} (middle)]. The H-H distances in these complexes vary from 1.6 \AA\ to 1.9 \AA, with the maximum ELF isovalue that captures their bonding being 0.62. This puts them in the covalent bonding class of hydrides \cite{belli2021strong}, which intriguingly has a spike in superconducting T$_c$ in materials with shortest H-H distance of 1.55 \AA (which all coincidentally are trihydrides, AH$_3$). To understand the electronic properties of this cell we plot the PDOS and find a small gap has opened, indicative of semimetallic behavior [Fig. \ref{fig:multicenter} (right)]. As DFT-PBE typically underestimates band gaps, this semimetallic phase may instead be semiconducting. The peak right above E$_F$ can be mostly attributed to the 7-H complexes. By looking at the integrated local density of states (ILDOS) from 0-0.5 eV above E$_F$ (Fig.\ \ref{fig:distort_ILDOS}(left)) we find that the QE projections underestimate the PDOS contribution of H$_s$ due to the significant spatial overlap of the Lu$_d$ orbital radius with the hydrogen atoms. The ILDOS indicates that the nature of the conducting network just above E$_F$ is that of the extended H complexes with high electron density weakly bonded to form a periodic metallic network by Lu$_d$ orbitals (Fig.\ \ref{fig:distort_ILDOS}(right)). 

The total DFT energy saved by this distortion from the high-symmetry SSCHA-relaxed structure amounts to over 22.5 meV/atom. However, we already saw the SSCHA converges to the high symmetry structure. This indicates that anharmonic stabilization from quantum effects in Fm$\overline{3}$m LuH$_{2.875}$N$_{0.125}$ is quite significant to prevent formation of bound hydrogen molecules, a well-known effect in metal hydrides \cite{wang2018large}.  

\section{$\alpha^2 F(\omega)$ Equations}
\label{sec:fig3}

The formulas for $\alpha^2 F(\omega)$ and that used for Fig. \ref{fig:SSCHA_results}(d) are:
\begin{align}
  & \gamma_\mu(\vec q) = \sum_{\vec k n m} \delta(\epsilon_{kn} - \epsilon_F) \delta(\epsilon_{k+q,n} - \epsilon_F) \abs{\sum_{a} \frac{\epsilon^a_\mu (\vec q)}{\sqrt{M_a}} d^a_{kn;k+qm}}^2 \\
  & \alpha^2 F(\omega) = \frac{1}{2\pi N_F}\sum_{q \nu} \frac{\gamma_{\mu}(q)}{\omega_{q\mu}} \delta(\omega - \omega_{q\mu}) \\
  & DOS^{A}_{weighted}(\omega) = \sum_{\mu q} M_\mu^0\frac{\abs{\epsilon^A_{\mu}(q)}^2}{M_A}\delta[\omega - \omega_\mu (q)]
\end{align}

Here $\gamma_\mu(\vec q)$ is the phonon lifetime of branch $\mu$ whose energy is $\omega_{q\mu}$, $\epsilon_F$ is the Fermi energy, $M_a$ is mass, $\epsilon^{a}_\mu (\vec q)$ indicates polarization,  $d^a_{kn;k+qm}$ are electron-phonon matrix elements, and N$_F$ is the density of states at the Fermi energy. 

 \begin{figure*}
   \centering
     \includegraphics[width=1.0\linewidth]{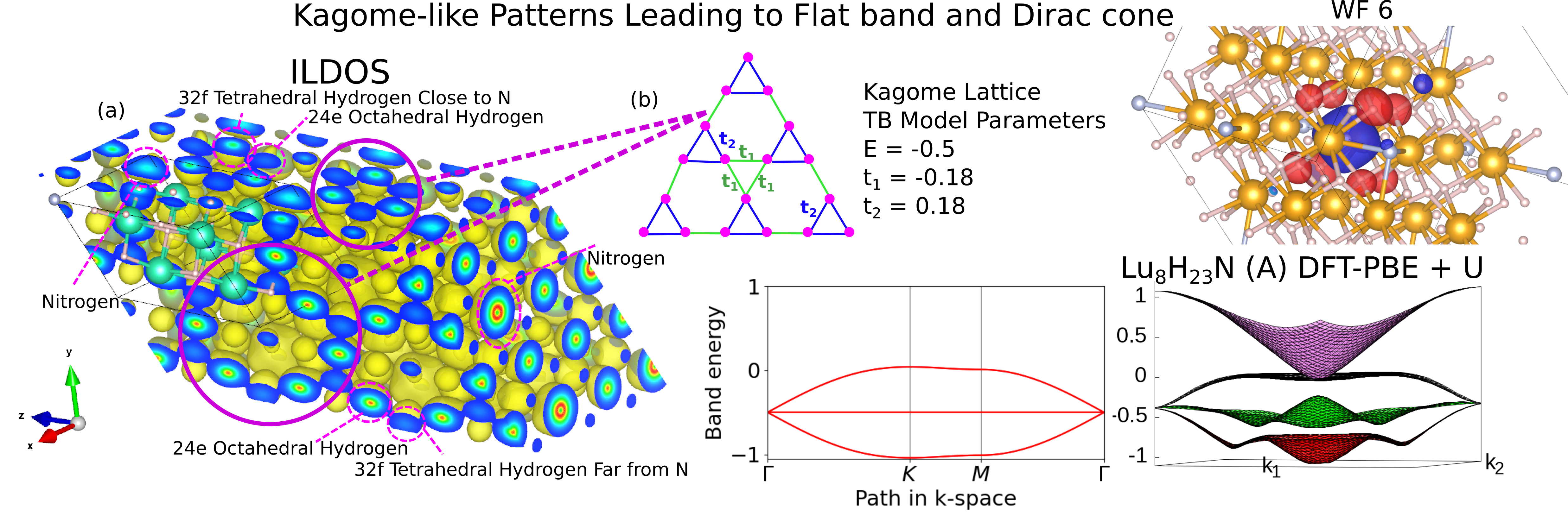}
     \caption{\justifying (a) ILDOS of Lu$_8$H$_{23}$N (A) from Ref. \cite{denchfield2024electronic}. Markers indicate Kagome-like patterns of the conduction states at E$_F$. (b) Model showing that an alternating-phase s-orbital tight-binding model on a Kagome lattice can lead to flat bands. (c) A wannier function from the minimal tight-binding model illustrating the alternating phases of the hydrogen s-orbitals in the lattice at E$_F$.}
   \label{fig:kagome}
 \end{figure*}

 \begin{figure*}
   \centering
     \includegraphics[width=1.0\linewidth]{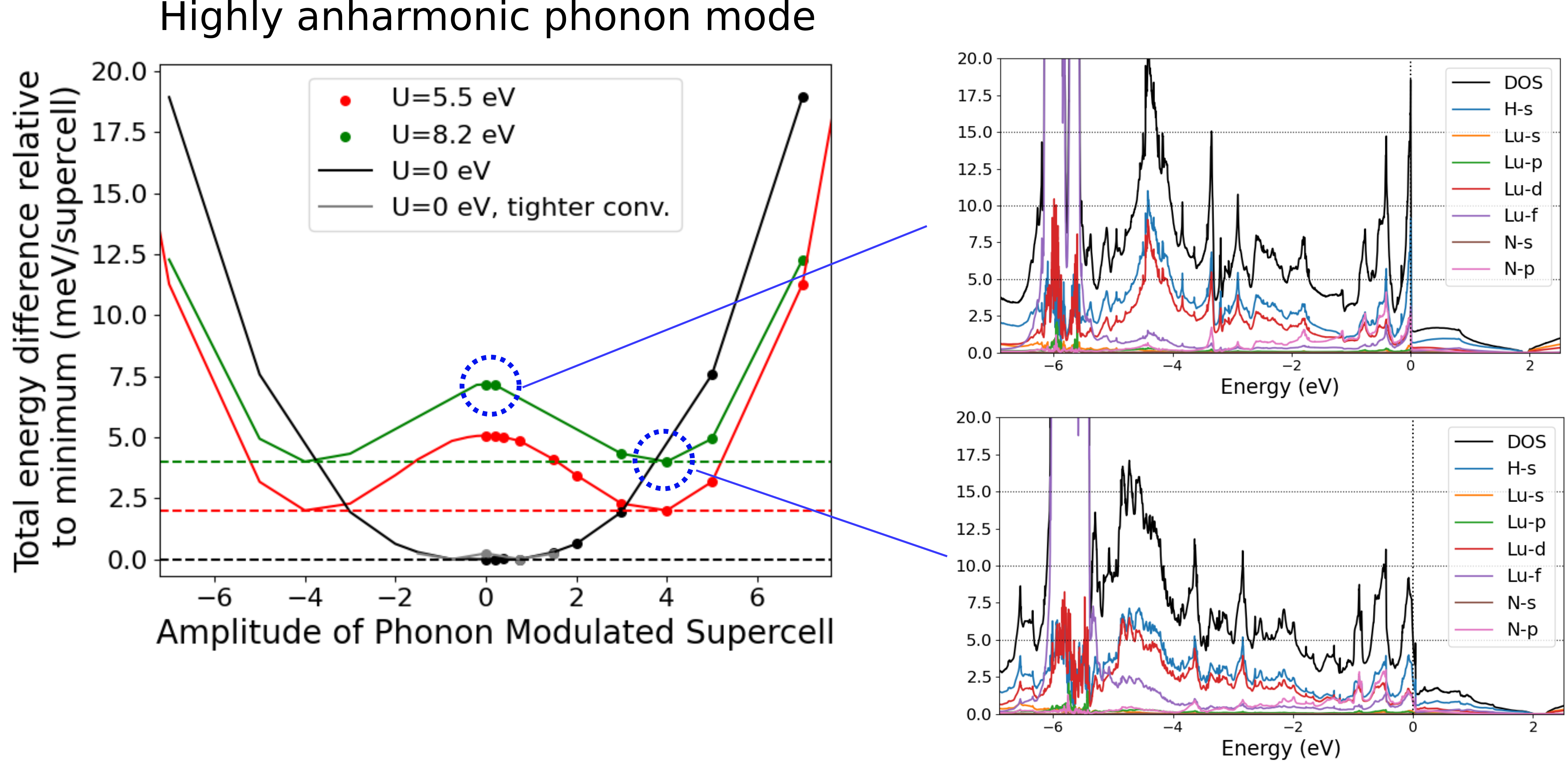}
     \caption{\justifying Anharmonicity analysis of Fig. \ref{fig:Lu8H23N_phonon_combined}(c) as well as the electronic PDOS of the system at 0 (a) and optimal (d) modulation, both using DFT+U with U on Lu$_d$ orbitals. The hydrogen states only move 5-10 meV below E$_F$ under the optimal modulation. The q=0 frequencies are displayed (b) and the anharmonicity of the PES along the mode with largest imaginary frequency is shown (c). The modulation is visually suggestive of superhydride units embedded in the crystal (e).}
   \label{fig:phonPDOS}
 \end{figure*}

 \begin{figure*}
   \centering
     \includegraphics[width=0.8\linewidth]{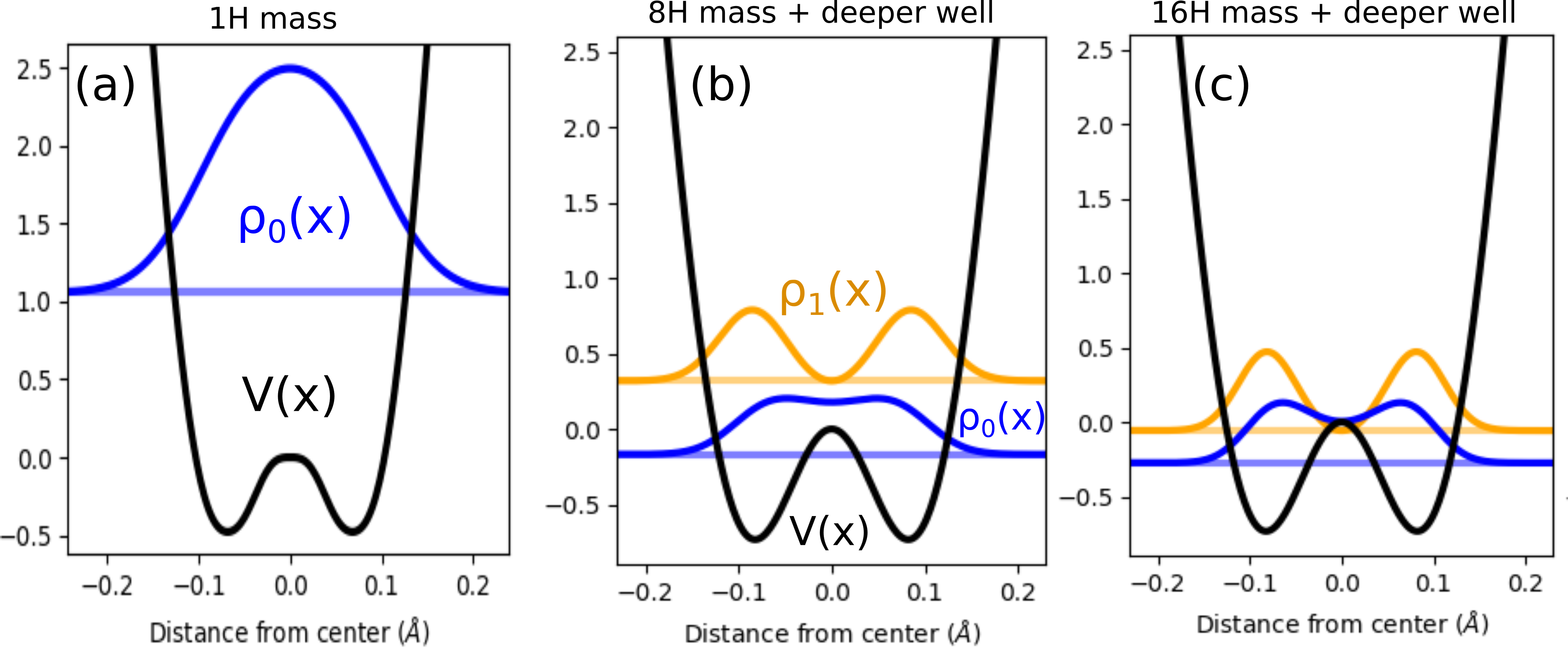}
     \caption{\justifying Solving the 1-particle Schrodinger equation for various V(x) (black curves), showing the ground state (blue) energy and density, as well as the first excited state density for higher masses (orange). The zero of density for each state is defined to be the energy of each state. (a) The ground state density when we attribute one hydrogen atomic mass to the particle with a PES similar to that in Fig. \ref{fig:Lu8H23N_phonon_combined}(c). (b) We deepen the well and increase the mass to eight hydrogens, flattening the ground state density. (c) We increase the mass to sixteen hydrogens which begins to become likely to symmetry break, however there is still significant density for the high-symmetry structure. The limited effect of particle mass has to do with the shallowness of the double well and the minima distances of 0.1 \AA\ per hydrogen. }
   \label{fig:phonModel}
 \end{figure*}

 \begin{figure*}
   \centering
     \includegraphics[width=1.0\linewidth]{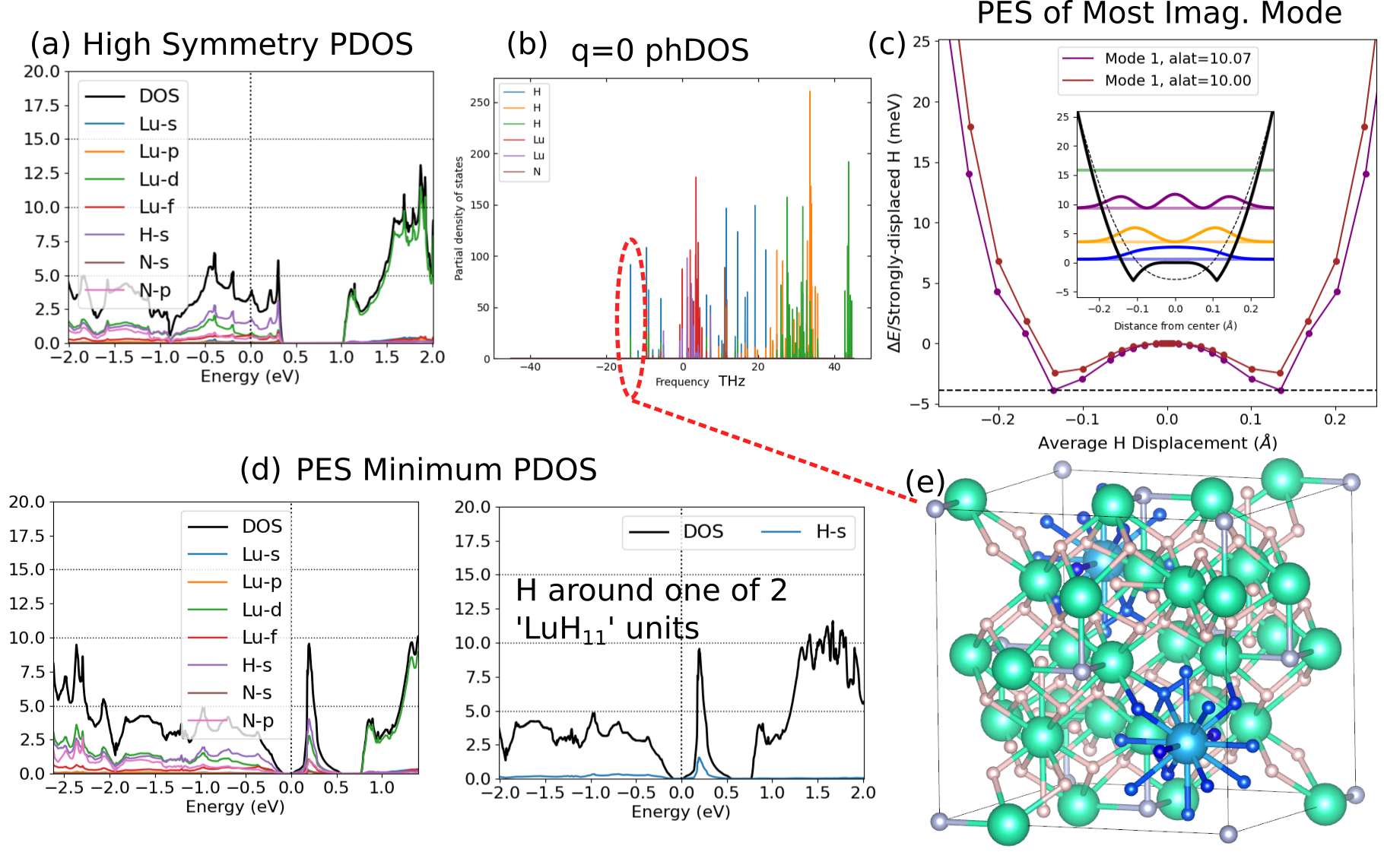}
     \caption{\justifying A similar anharmonicity analysis of the phonon indicating a classical structural instability in LuH$_{2.75}$N$_{0.125}$. The PDOS of both high symmetry and perturbed structures is shown, as well as a visualization of the distortion on the lower right. }
   \label{fig:LuH22N_phon}
 \end{figure*}
 
 \begin{figure*}
   \centering
     \includegraphics[width=1.0\linewidth]{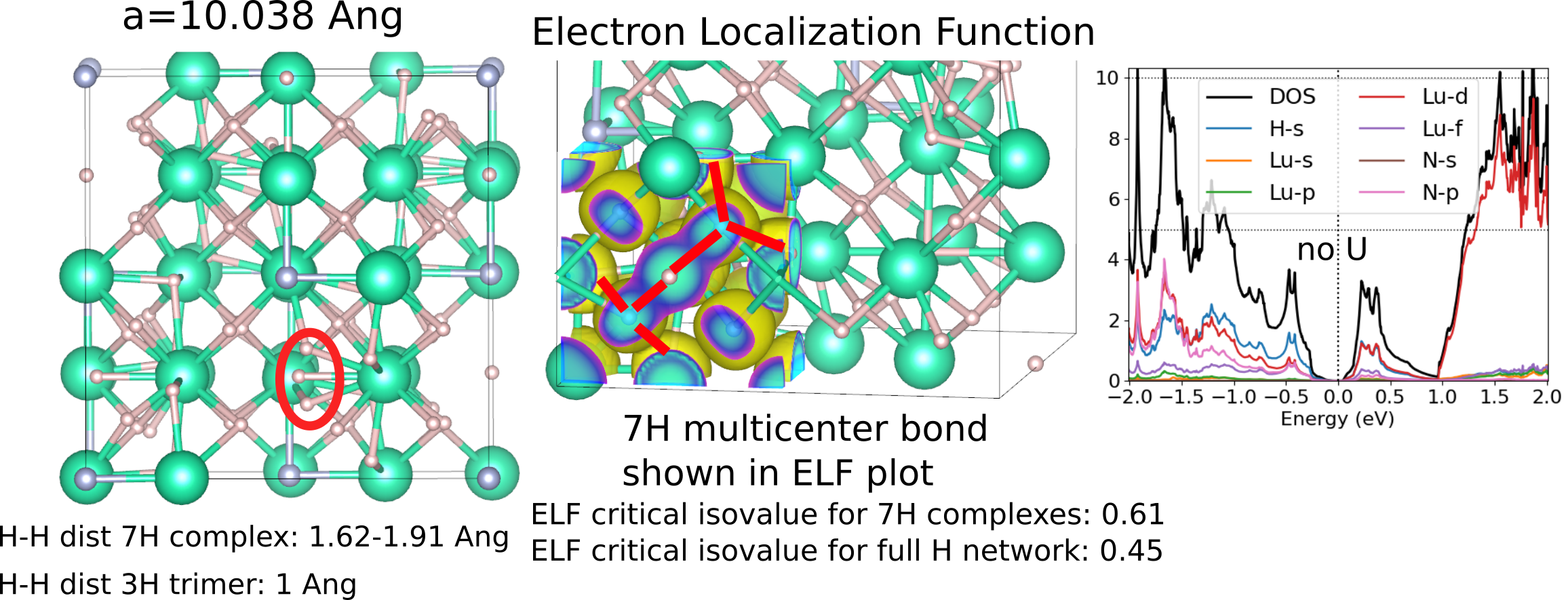}
     \caption{\justifying The LuH$_{2.875}$N$_{0.125}$ structure classically relaxed from the SSCHA-relaxed starting point (left), with a red circle indicating a hydrogen trimer. The corresponding PDOS computed at the DFT-PBE level shows a semiconducting or semimetallic state, with a portion of the PDOS above E$_F$. The Electron-Localization-Function illustrates one of two 7H multicenter bonds formed in the unit cell (right), with red lines marking the network of the bonded complex.}
   \label{fig:multicenter}
 \end{figure*}

 \begin{figure*}
   \centering
     \includegraphics[width=1.0\linewidth]{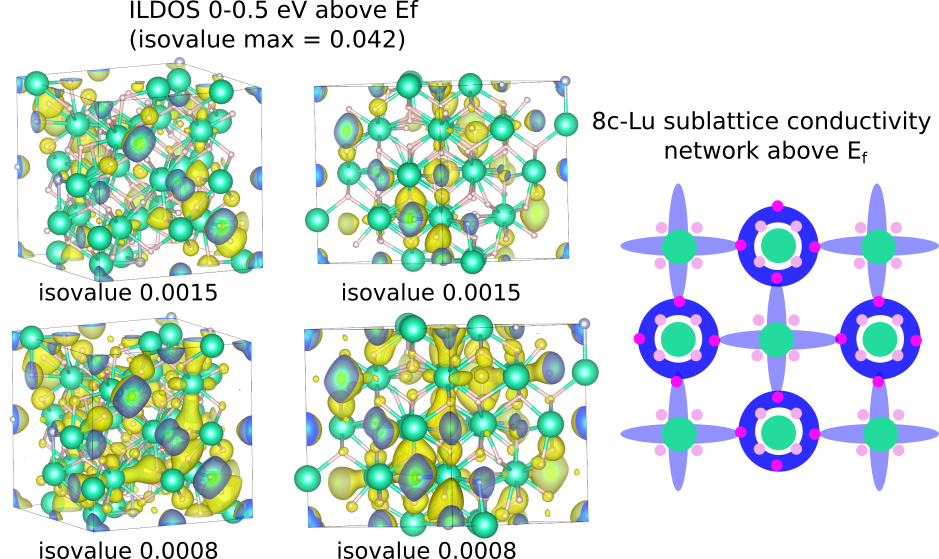}
     \caption{\justifying Integrated local density of states (ILDOS) for the classically distorted Fm$\overline{3}$m LuH$_{2.875}$N$_{0.125}$ 0-0.5 eV above E$_f$ (see Fig. \ref{fig:multicenter}) at a variety of different isovalues. We can see the majority of the high electron density regions are focused on hydrogen, especially in the 7H complexes, with a nontrivial amount around the nitrogen atoms. There are low electron density regions corresponding to Lu$_d$ orbitals that weakly couple the 7H complexes together. We visualize the conduction pathway of the PDOS peak just above E$_F$ (Fig. \ref{fig:multicenter}) of the 8c-Lu sublattice alongside the hydrogen surrounding them on the right. Light and dark blue indicate low and high electron density. Teal atoms are lutetium, with light and dark pink denoting tetrahedral and octahedral hydrogens respectively.}
   \label{fig:distort_ILDOS}
 \end{figure*}

  \begin{figure}
   \centering
     \includegraphics[width=1.0\linewidth]{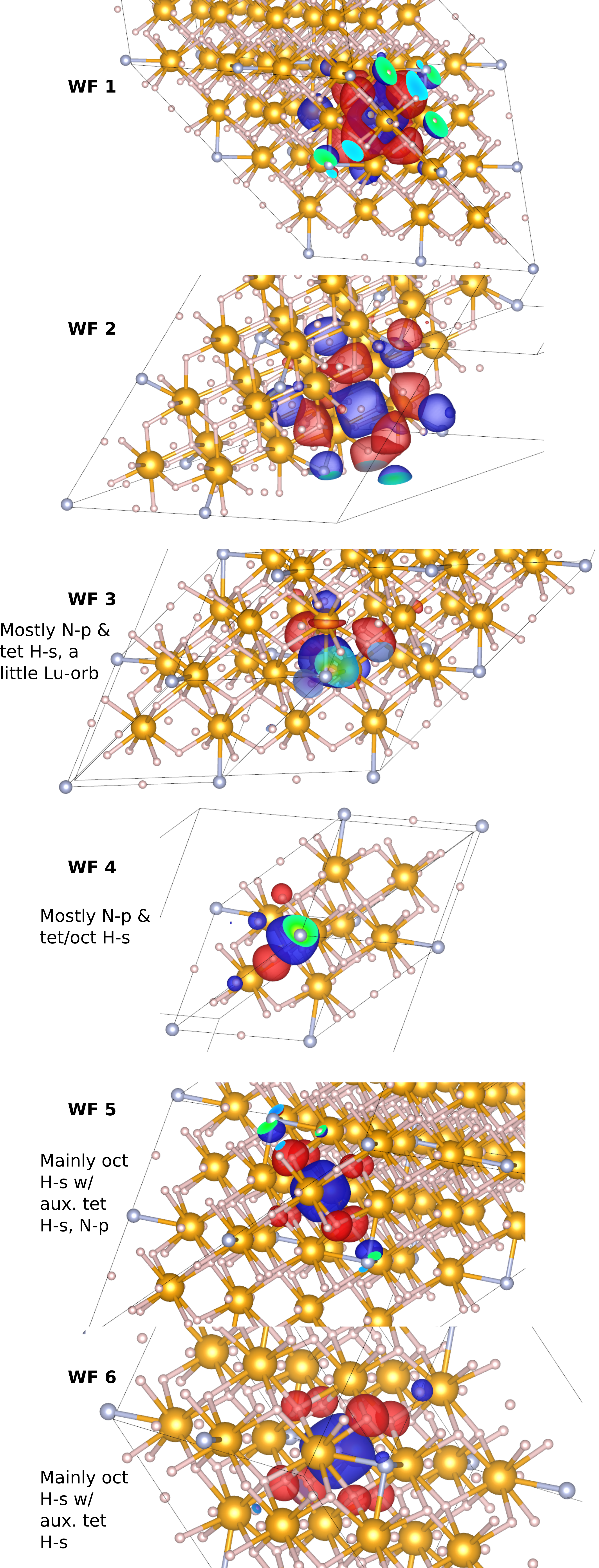}
     \caption{\justifying The 6 Wannier functions for the minimal tight-binding model for bands crossing E$_F$.}
   \label{fig:WFs}
 \end{figure}

  \begin{figure}
   \centering
     \includegraphics[width=1.0\linewidth]{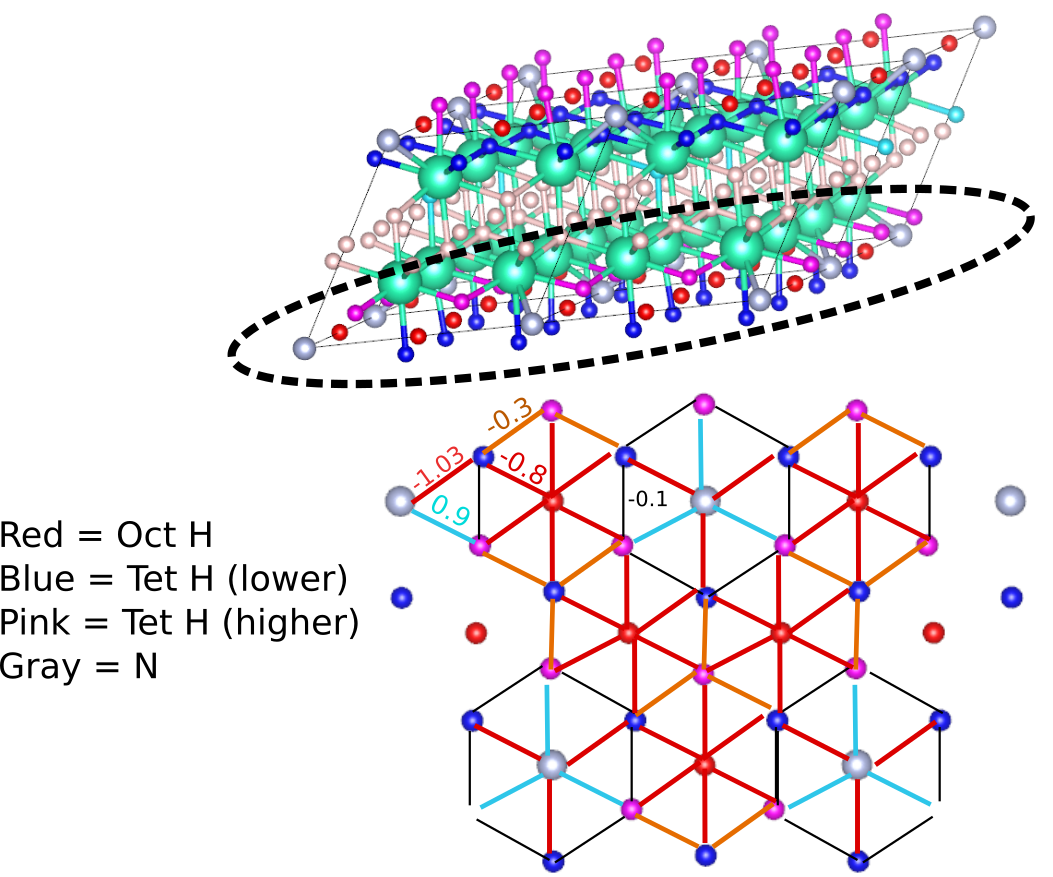}
     \caption{\justifying Illustration of the nitrogen and hydrogen sublattice as well as the hoppings in a TB model of atomic Wannier functions. Color indicates sign, and alternating signs can be shown to be induced around the nitrogen atoms, a likely signature of quantum interference effects which can lead to flat bands.}
   \label{fig:WFs_26orb}
 \end{figure}

  \begin{figure}
   \centering
     \includegraphics[width=1.0\linewidth]{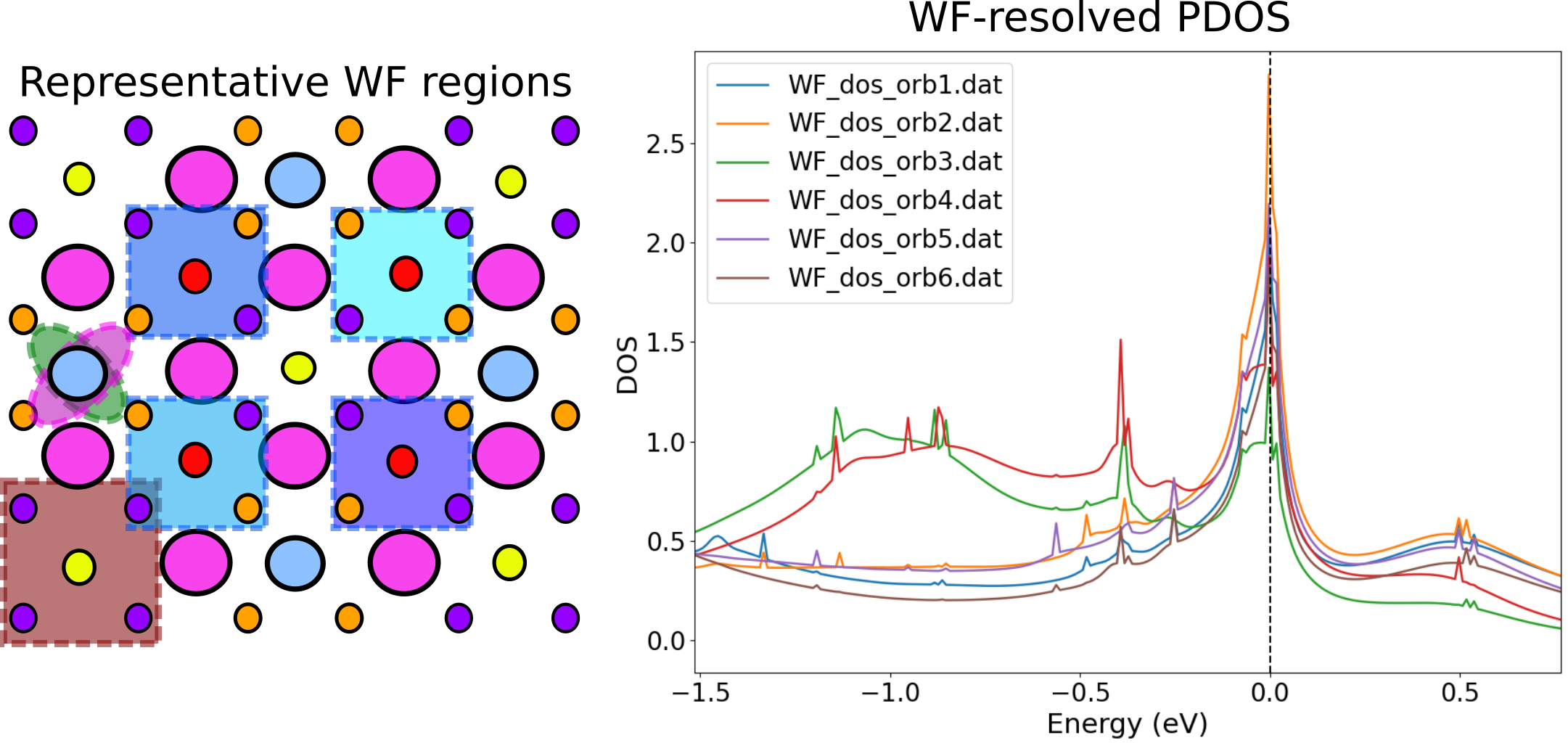}
     \caption{\justifying Cartoon of the relative centers of the 6 Wannier functions for the minimal tight-binding model (left), as well as the WF-resolved PDOS (right).}
   \label{fig:WFs_pdos}
 \end{figure}
 
 \newpage
 \section{6-orbital Tight-Binding Model}
 \label{sec:tb_elements}
 Here we list the tight-binding matrix elements of the 6-WF model described in the text. We only list the elements with matrix elements above 0.1 in absolute value; we are happy to share the remaining elements upon request. Rx,Ry,Rz refer to unit cell indices, m and n refer to WF indices, and H refers to the tight-binding elements.
\begin{verbatim}
   Rx   Ry   Rz    m    n   H[Re] (eV)   H[Im] (eV)
   -2    0    0    5    3   -0.235800    0.000933
   -1   -1    0    5    3    0.228939   -0.000905
   -1   -1    0    5    6   -0.221397   -0.000087
   -1    0   -1    1    4    0.459268   -0.000066
   -1    0   -1    1    6   -0.342960   -0.000803
   -1    0    0    5    2    0.430709   -0.001180
   -1    0    0    6    3    0.443425   -0.001939
   -1    0    0    1    6   -0.343030   -0.000803
   -1    0    0    2    6   -0.364583   -0.001143
   -1    0    0    5    6   -0.211822   -0.000083
   -1    1    0    1    1    0.202198   -0.000000
    0   -1    0    2    6   -0.349415   -0.001095
    0   -1    1    5    1    0.401877   -0.000782
    0   -1    1    2    2    0.229908   -0.000000
    0    0   -2    1    4   -0.471485    0.000067
    0    0   -1    2    2    0.228012    0.000000
    0    0    0    1    1   14.020245   -0.000000
    0    0    0    5    1    0.380270   -0.000740
    0    0    0    2    2   14.222420   -0.000000
    0    0    0    5    2    0.402105   -0.001102
    0    0    0    3    3   14.007575    0.000000
    0    0    0    6    3   -0.437418    0.001912
    0    0    0    4    4   14.234139    0.000000
    0    0    0    1    5    0.380270    0.000740
    0    0    0    2    5    0.402105    0.001102
    0    0    0    5    5   14.079154    0.000000
    0    0    0    3    6   -0.437418   -0.001912
    0    0    0    6    6   13.838899    0.000000
    0    0    1    2    2    0.228012   -0.000000
    0    0    2    4    1   -0.471485   -0.000067
    0    1   -1    2    2    0.229908    0.000000
    0    1   -1    1    5    0.401877    0.000782
    0    1    0    6    2   -0.349415    0.001095
    1   -1    0    1    1    0.202198    0.000000
    1    0    0    6    1   -0.343030    0.000803
    1    0    0    6    2   -0.364583    0.001143
    1    0    0    2    5    0.430709    0.001180
    1    0    0    6    5   -0.211822    0.000083
    1    0    0    3    6    0.443425    0.001939
    1    0    1    4    1    0.459268    0.000066
    1    0    1    6    1   -0.342960    0.000803
    1    1    0    3    5    0.228939    0.000905
    1    1    0    6    5   -0.221397    0.000087
    2    0    0    3    5   -0.235800   -0.000933
\end{verbatim}
 
\begin{figure}
   \centering
     \includegraphics[width=1.0\linewidth]{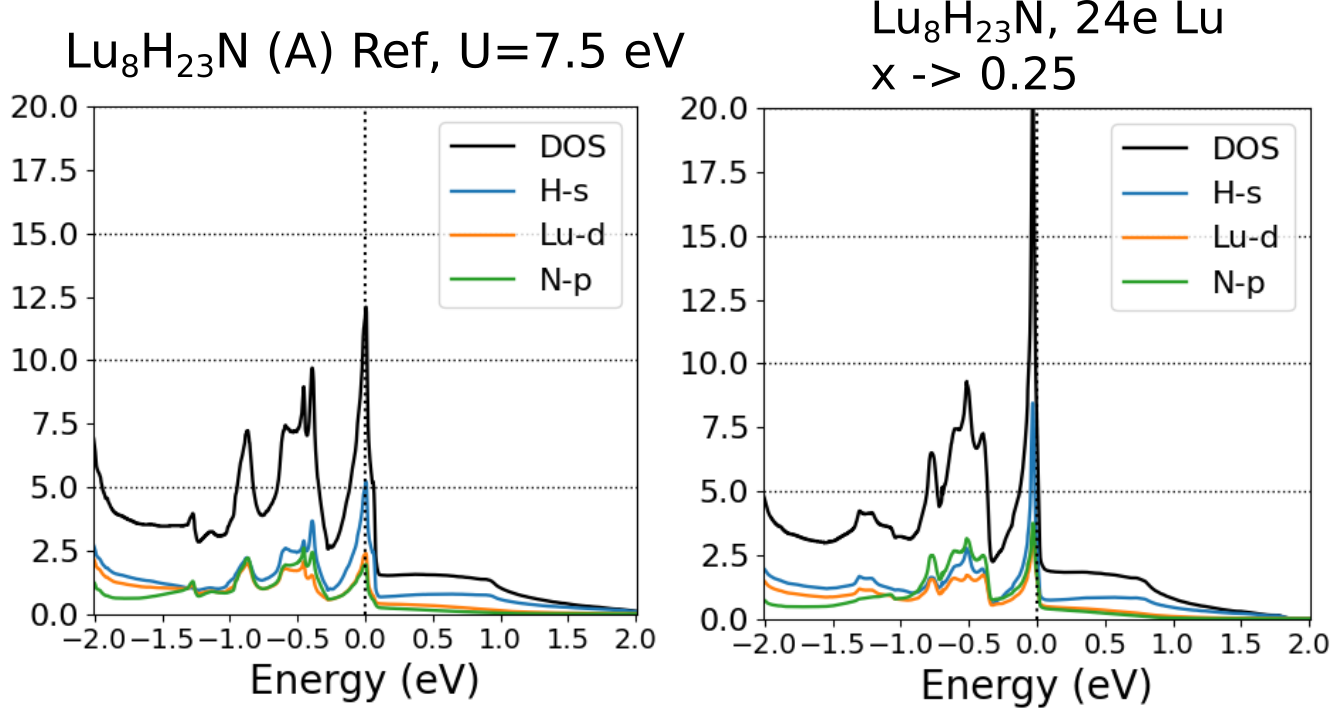}
   \caption{\justifying PDOS computed with DFT+U of Lu$_8$H$_{23}$N (A) at the PBE-relaxed coordinates (left) and when moving the 24e-Lu back to the symmetric x=0.25 position.}
   \label{fig:Lu_sym}
 \end{figure}

 \begin{figure}
   \centering
     \includegraphics[width=1.0\linewidth]{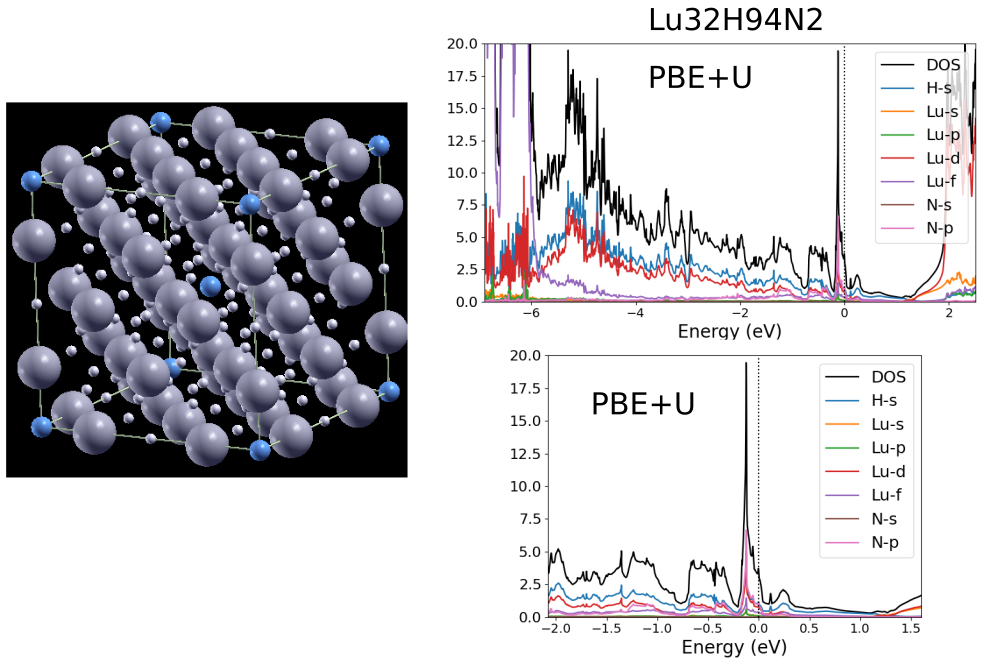}
   \caption{\justifying PDOS of Lu$_{16}$H$_{47}$N with DFT+U, U=7.5 eV. There is a flat band under E$_F$ of nitrogen character which persists upon relaxation of the unit cell.}
   \label{fig:Lu16H47N}
 \end{figure}

\end{document}